\documentclass[useAMS,usenatbib,usegraphicx]{mn2e}
\usepackage{times}
\newcommand{\hi}{\mbox{H{\sc i}}}
\newcommand{\hii}{\mbox{H{\textsc{ii}}}}
\newcommand{\ha}{H$\alpha$}
\newcommand{\pa}{$P.A.$}
\newcommand{\nii}{\mbox{[N{\textsc{ii}}]}}
\newcommand{\kmps}{km s$^{-1}$}

\newcommand{\as}[2]{$#1''\,\hspace{-1.7mm}.\hspace{.1mm}#2$}
\newcommand{\am}[2]{$#1'\,\hspace{-1.7mm}.\hspace{.0mm}#2$}
\newcommand{\amd}[2]{$#1\degr\,\hspace{-1.7mm}.\hspace{.0mm}#2$}
\newcommand{\coa}[3]{#1$^h$#2$^m$#3$^s$}
\newcommand{\cod}[3]{#1\degr #2\arcmin #3\arcsec}
\newcommand{\fov}{f.-o.-v.}

\newcommand{\fantomm}{{\texttt {\textsc{FaNTOmM}}}}
\newcommand{\cigale}{{\texttt {\textsc{Cigale}}}}
\newcommand{\sauron}{{\texttt {\textsc{SAURON}}}}

\title[A Virgo high-resolution \ha\ kinematical survey :~II. The Atlas]
{A Virgo high-resolution \ha\ kinematical survey :~II. The Atlas\thanks{E-mail: chemin@astro.umontreal.ca -- Based on observations collected at the 
European Southern Observatory (program 69.B-0496),  the Observatoire du mont M\'egantic (Qu\'ebec, Canada),  
 the Observatoire de Haute-Provence (France) and the Canada-France-Hawaii  Telescope, which is operated by the National Research Council of Canada, the Centre
National de la Recherche Scientifique de France and the University of Hawaii.}}
\author[L. Chemin et al.] {L. Chemin$^{1,2}$,  C. Balkowski$^{2,5}$,
V. Cayatte$^{3,5}$,  C. Carignan$^{1,5}$, P. Amram$^{4,5}$, O. Garrido$^{2,4}$, 
\newauthor O. Hernandez$^{1,4,5}$, M. Marcelin$^{4,5}$, C. Adami$^{4}$, A. Boselli$^{4}$,  J. Boulesteix$^{4}$\\
$^1$D\'epartement de Physique and Observatoire du mont M\'egantic, Universit\'e de Montr\'eal, C.P. 6128, Succ. centre-ville, Montr\'eal, Qc, Canada, H3C 3J7\\
$^2$Observatoire de Paris, section Meudon, GEPI, CNRS-UMR 8111 \& Universit\'e Paris 7, 5 Pl. Janssen, 92195 Meudon, France\\
$^3$Observatoire de Paris, section Meudon, LUTH, CNRS-UMR 8102 \& Universit\'e Paris 7, 5 Pl. Janssen, 92195 Meudon, France\\
$^4$Observatoire Astronomique de Marseille Provence, 2 Pl. Le Verrier, 13248 Marseille, France\\
$^5$ Visiting Astronomer, Canada--France--Hawaii Telescope.}

\pagerange{\pageref{firstpage}--\pageref{lastpage}} \pubyear{2005}

\begin{document}
\label{firstpage}
\maketitle

\begin{abstract}
A catalog of ionized gas velocity fields for 
a sample of 30  spiral and irregular galaxies of the Virgo cluster has been obtained by using three-dimensional optical data. 
The aim of this survey is to study the influence of high density environments on the gaseous 
kinematics of local cluster galaxies. 
Observations of the \ha\ line by means of Fabry-Perot interferometry have been performed  
at the Canada-France-Hawaii, ESO 3.6m, Observatoire de Haute-Provence 
1.93m and Observatoire du mont M\'egantic telescopes at angular and spectral samplings from 
 \as{0}{4} to \as{1}{6} and 7 to 16 \kmps.  
A recently developed, automatic and adaptive spatial binning technique is used to reach a nearly constant 
signal-to-noise ratio ($S/N$) over the whole field-of-view, allowing to keep a high spatial resolution in high $S/N$ 
regions and extend the detection of signal in low $S/N$ regions. This paper is part of a series and presents 
the integrated emission-line and velocity maps of the galaxies. Both \ha\ morphologies and kinematics 
exhibit signs of perturbations in the form of e.g. external filaments, inner and nuclear spiral- and ring-like structures, 
inner kinematical twists, kinematical decoupling of a nuclear spiral, streaming motions along spiral arms and   
misalignment between kinematical and photometric orientation axes.  
 
 \end{abstract}

\section{Introduction}
\label{sec:introduction}

Dense environments such as the cores of  galaxy clusters are known to give to disc galaxies  
photometric and spectral  properties that may considerably differ from
those of galaxies found in lower density environments, like the clusters periphery or the field. 

Manifestations of environmental effects in local galaxy clusters  are  illustrated
by  the morphology-density relation (Dressler 1980;
Binggeli, Tammann \& Sandage 1987), the neutral hydrogen deficiency of spiral galaxies 
 in the cluster core (Chamaraux, Balkowski \& G\'erard  1980; 
  Giovanelli \& Haynes 1985; Gavazzi 1987; Cayatte et al. 1990;  Bravo-Alfaro et al. 2000) but also 
at large clustercentric distances (Solanes et al. 2001; but see Sanchis et al. 2004) or the occurence of peculiar \ha\ morphologies   
for spirals within the highest galaxy surface density  regions (Moss \& Whittle 2000; Koopmann \& Kenney 2004; Vogt et al. 2004).
Centres of nearby clusters are dominated by spheroidals, gas-poor, non star-forming discs (lenticulars) and ellipticals while
lower density regions are gas-rich, spiral discs dominated. 
Many works  have furthermore brought evidence that the properties of cluster spirals 
strongly evolve with the redshift (see Poggianti 2004 and references therein).  
This can be seen in the evolution of the morphology-density relation  (Dressler et al. 1997;  Fasano et al. 2000; Smith et al. 2005) 
and of galaxy colours (Butcher \& Oemler 1978, 1984). 
Distant,  intermediate redshifts ($0.2 \la z \la 0.8$)  clusters are 
populated by a larger number of star-forming spirals rather than by passive lenticulars. 
These results rise the question of a possible morphological transformation of spirals into 
S0s since a redshift of $z \sim 0.5$ and of the evolution of the star formation activity of galaxies 
(e.g. Couch \& Sharples 1987; Balogh et al. 1999; Dressler et al. 1999; Adami et al.  2000). 
An active star formation period has been rapidly quenched (e.g. Poggianti et al. 1999), leaving the 
star-forming galaxies into a poststarburst phase, a probably important step during their possible evolution 
towards early-type objects (e.g. Tran et al. 2003). 
The population of massive elliptical galaxies does not  vary as a function of redshift  
(Dressler et al. 1997; Postman  et al. 2005; Smith  et al. 2005) 
and  their formation appears to go back to 
the earliest epochs ($z \ga 2-3$, e.g. Ellis et al. 1997; van Dokkum et al. 1998; van Dokkum \& Franx 2001).

 The recent theoretical works armed with N-body or hydrodynamical  numerical simulations have furthermore considerably improved our understanding 
 of the physical processes  that could drive the evolutions of the stellar populations 
 and the morphological properties of cluster galaxies. Among several theories, two  preferred invoked processes are tidal 
 effects and ram pressure stripping (RPS).  On one hand, simulations of tidal effects can account for a slow morphological change within a time-scale of few Gyrs
  when a galaxy enters a cluster and undergoes several gravitational interactions with other galaxies and with the cluster
  tidal field. Instances of such models are the ``galaxy  harassment"  (Moore at al. 1996, 1998, 1999)   
  or the ``tidal heating" (Gnedin 2003a,b). In these models, low density galaxies such as low surface brightness discs  
lose almost all their stars; their remnants appear to become dwarf spheroidal systems and their stripped stars the diffuse intracluster light.
Galaxy mergers could participate to the formation of central giant ellipticals and 
cD galaxies during the earliest phase of the cluster formation (Merritt 1985;  Dubinski 1998). 
On another hand, ram pressure stripping (Gunn \& Gott 1972) could explain the \hi\ deficiency of today spirals, and perhaps 
why the  gas reservoir has been emptied and the star formation stopped in distant spirals. 
Here the hot gas of the intra-cluster medium (ICM) exerts a pressure
on the gaseous disc of a galaxy, acting like a wind that can efficiently sweep it in several tens of Myrs (Abadi, Moore \& Bower 1999; Quilis, 
Moore \& Bower 2000; Schulz \& Struck 2001; Vollmer et al. 2001; Roediger \& Hensler 2005). 
Eccentric orbits allow galaxies to go deeper through the cluster core and thus to be more efficently stripped (Vollmer et al. 2001) 
and could explain the presence of \hi\ deficient spirals observed at large cluster-centric radius (Solanes et al. 2001).  
Because this mechanism does not involve gravitational forces, at a first order it cannot influence  the (old) stellar population of a disc. 
At a second order, some simulations claim that the rise of gas surface density when a disc passes close 
to the cluster core and the gas shocks induced by the return of expelled gas 
into a parent galaxy can enhance the  formation of stars (e.g. Vollmer et al. 2001). 
This is nonetheless probably not sufficient to explain the observed morphological segregation and 
the two above-mentionned mechanisms may be both at work to explain the observations (Poggianti et al. 1999). 

The kinematical and dynamical properties of cluster galaxies have also 
been the subject of several works although they have been less studied than their spectro-photometric ones. 
In local clusters, optical and \hi\ rotation curves are used to study the Tully-Fisher relation of spirals 
(e.g. Schommer et al. 1993; Dale et al. 1997, 1999; Verheijen  2001) or the shape of the curves and its connection with 
the environment (e.g. Amram et al. 1993, 1996; Sperandio et al. 1995; Adami et al. 1999; 
Rubin et al. 1999; Dale et al. 2001). Rubin et al. (1999) used a large sample of optical rotation curves of Virgo cluster galaxies and 
found that the curves harbouring the most disturbed shapes have a distribution of systemic 
velocities very well correlated with  that of the Virgo ellipticals and are more and more frequent (relative to the regular rotation curves)
as the cluster-centric radius from M87 (at the cluster core) increases. 
Both these results suggest highly eccentric orbits for spirals (see also Adami, Biviano \& Mazure 1998). Moreover, the perturbed curves
are not found preferentially in \hi\ deficient galaxies, suggesting that the two mechanisms responsible for the deficiency and the observed kinematical disturbances differ. 
In distant clusters, observations of galaxies currently concentrate on the study of their Tully-Fisher relation   
 (e.g. Ziegler et al. 2003; Bamford et al. 2005) and it is still too early to learn about any evolution as a function of redshift 
 of their kinematical disturbances.
    
In this work, we are interested in studying the kinematical and dynamical properties 
of Virgo spiral galaxies. For that purpose a  sample of 30 galaxies obtained using 
 optical data at high angular and spectral resolutions has been observed.  
Typical questions that are still to be answered and that motivate our study 
are~: how do the properties of gaseous velocity fields, their asymmetries (induced by e.g. a warp, 
a lopsidedness, a bar or spiral pattern) and of dark matter haloes depend on such a dense environment ? 
Do morphologically disturbed spirals exhibit kinematical perturbations ?
How does the kinematics correlate with the star formation efficiency ? 
 Long-slit spectroscopy data can hardly answer these questions because they only 
inform us about the kinematics along a few directions (mainly the photometric major axis) and are not appropriate 
to investigate the full kinematical structure of spiral galaxies. 
The recent advent of a new generation of 3D optical spectroscopic instruments 
 offers very promising perpectives for the stellar and/or gaseous kinematics of  large samples of nearby galaxies 
 spanning  large ranges of luminosity and morphology (de Zeeuw et al. 2002;  Emsellem et al. 2004; Bershady et al. 2005; 
 Daigle et al. 2005a; Hernandez et al. 2005a). These instruments 
 are e.g. fiber-fed or lenslet array integral-field spectrographs (Bacon et al. 2001; Bershady et al. 2004) 
 or Fabry-Perot (FP) spectro-imagers (Gach et al. 2002).  We want to extend the available data with a sample of  
 gaseous velocity fields of Virgo cluster spirals using the Fabry-Perot interferometry technique.  
 3D optical data are a good compromise for the study of environmental effects in clusters  
 due to a better distribution and resolution of the gas kinematics than for radio data.  
 Virgo cluster spirals are indeed  \hi\ deficient, often showing \hi\ discs smaller than the optical ones (Cayatte et al. 1990). 
 Moreover  the neutral hydrogen often lack their central parts, regions where the molecular gas is mainly present (Sofue et al. 2003a,b).  
 As an additional motivation to our study, the 3D observations of gas  will be compared 
 to results of N-body and/or hydrodynamical simulations, as already done for a few Virgo galaxies (see e.g. Vollmer et al. 2004). 
 It could help to constrain  the evolutionary scenario of peculiar objects in the cluster (e.g. ram pressure stripping, encounters of galaxies).  
 
 This article is the second from a series. The first paper was dedicated to NGC 4438, a
  prototype of a galaxy interacting with the ICM and with companions (Chemin et al. 2005). 
  This article presents a catalog of 2D \ha\ emission-line and kinematical maps for 30 Virgo galaxies observed 
  with the \cigale\ and \fantomm\footnote{Fabry-Perot de Nouvelle Technologie de l'Observatoire du mont M\'egantic. See http://www.astro.umontreal.ca/fantomm} ~instruments.  
  The article presents the sample of the galaxies, the principles of the data reduction and of the adaptive binning in sections~\ref{obs} and~\ref{reduc}. Sections~\ref{results} 
and~\ref{analysis} briefly present the results and a discussion. Section~\ref{conclusions} resumes the results presented in the article. 
Comments for the individual galaxies are given in Appendix~\ref{comments} and their kinematics 
is presented in Appendix~\ref{velmap}. 

 The data will be soon available from a Fabry-Perot data-base. 
 They can be obtained  from the authors before upon request. 

\section{Definition of the sample}
\label{obs}

\begin{table*}
\centering
\begin{minipage}{168mm}
    \caption{Virgo galaxy targets parameters}
    \label{targchar}
    \begin{tabular}{ccccclrcccc}
\hline\hline
VCC & NGC-IC & Messier & $\alpha_{2000}$ & $\delta_{2000}$  & Type   & $B^0_t$ &
 $v_{\rm sys}$ & Dist. to M87 & Memb. & \hi\ deficiency  \\
(1) & (2)    & (3) & (4)             & (5)             & (6)  & (7) & (8) &  (9)    &   (10)	  	& (11)      \\
\hline																	    
0089 & NGC 4189  &  & \coa{12}{13}{47.3} & \cod{+13}{25}{29}     & SAB(rs)cd?	     & 12.21  & 2115  & \amd{4}{4}  & M & -0.03   \\ 
0157 & NGC 4212  &  & \coa{12}{15}{39.3} & \cod{+13}{54}{05}     & SAc:      	     & 11.35  & -81   & \amd{4}{1}  & N & 0.61   \\ 
0226 & NGC 4237  &  & \coa{12}{17}{09.4} & \cod{+15}{19}{33}     & SAB(rs)bc 	     & 12.18  & 867   & \amd{4}{5}  & N & 0.48   \\ 
0307 & NGC 4254  & M99 & \coa{12}{18}{49.6} & \cod{+14}{24}{59}  & SA(s)c    	     & 10.10  & 2407  & \amd{3}{6}  & N & 0.01  \\ 
0465 & NGC 4294  &  & \coa{12}{21}{17.8} & \cod{+11}{30}{40}     & SB(s)cd   	     & 11.83  & 359   & \amd{2}{5}  & N & 0.10  \\ 
0483 & NGC 4298  &  & \coa{12}{21}{32.8} & \cod{+14}{36}{22}     & SA(rs)c   	     & 11.62  & 1135  & \amd{3}{2}  & A & 0.34  \\ 
0567 &  IC 3225  &  & \coa{12}{22}{39.1} & \cod{+06}{40}{42}     & Sdm:      	     & 13.87  & 2363  & \amd{6}{1}  & B & 0.36    \\	  
0596 & NGC 4321  & M100 & \coa{12}{22}{54.9} & \cod{+15}{49}{21} & SAB(s)bc  	     & 9.98   & 1571  & \amd{4}{0}  & A & 0.53  \\ 	  
0692 & NGC 4351  &  & \coa{12}{24}{01.5} & \cod{+12}{12}{18}     & SB(rs)ab pec:     & 12.79  & 2310  & \amd{1}{7}  & A & 0.66	\\	  
0874 & NGC 4405  &  & \coa{12}{26}{07.1} & \cod{+16}{10}{51}     & SA(rs)0/a:	     & 12.83  & 1747  & \amd{4}{0}  & A & 0.95 	\\  
1043 & NGC 4438  &  & \coa{12}{27}{45.6} & \cod{+13}{00}{32}     & SA(s)0/a pec:     & 10.49  & 71    & \amd{1}{0}  & A & 1.33$^a$   \\
1110 & NGC 4450  &  & \coa{12}{28}{29.6} & \cod{+17}{05}{06}     & SA(s)ab	     & 10.75  & 1954  & \amd{4}{7}  & A & 0.95    \\
1145 & NGC 4457  &  & \coa{12}{28}{59.0} & \cod{+03}{34}{14}     & (R)SAB(s)0/a      & 11.55  & 882   & \amd{8}{8}  & S & 0.86     \\
1379 & NGC 4498  &  & \coa{12}{31}{39.5} & \cod{+16}{51}{10}     & SAB(s)d	     & 12.33  & 1507  & \amd{4}{5}  & A & 0.15  \\
1401 & NGC 4501  & M88 & \coa{12}{31}{59.2} & \cod{+14}{25}{14}  & SA(rs)b	     & 9.86   & 2281  & \amd{2}{1}  & A & 0.55  \\
1508 & NGC 4519  &  & \coa{12}{33}{30.3} & \cod{+08}{39}{16}     & SB(rs)d	     & 12.15  & 1220  & \amd{3}{8}  & S & -0.26  \\
1554 & NGC 4532  &  & \coa{12}{34}{19.3} & \cod{+06}{28}{04}     & IBm  	     & 11.92  & 2012  & \amd{6}{0}  & S & -0.37  \\
1555 & NGC 4535  &  & \coa{12}{34}{20.3} & \cod{+08}{11}{52}     & SAB(s)c	     & 10.32  & 1961  & \amd{4}{3}  & S & 0.19   \\ 
1562 & NGC 4536  &  & \coa{12}{34}{27.1} & \cod{+02}{11}{16}     & SAB(rs)bc	     & 10.58  & 1808  & \amd{10}{2} & S & 0.15  \\ 
1615 & NGC 4548  & M91 & \coa{12}{35}{26.4} & \cod{+14}{29}{47}  & SBb(rs)	     & 10.79  & 486   & \amd{2}{4}  & A & 0.80	\\ 
1673 & NGC 4567  &  & \coa{12}{36}{32.7} & \cod{+11}{15}{28}     & SA(rs)bc	     & 11.79  & 2274  & \amd{1}{8}  & A & 0.43	\\ 
1676 & NGC 4568  &  & \coa{12}{36}{34.3} & \cod{+11}{14}{19}     & SA(rs)bc	     & 11.18  & 2255  & \amd{1}{8}  & A & 0.58	\\ 
1686 &  IC 3583  &  & \coa{12}{36}{43.5} & \cod{+13}{15}{34}     & Sm		     & 13.24  & 1121  & \amd{1}{7}  & A & 0.79   \\ 
1690 & NGC 4569  & M90 & \coa{12}{36}{49.8} & \cod{+13}{09}{46}  & SAB(rs)ab	     & 9.79   & -235  & \amd{1}{7}  & A & 1.07	\\ 
1696 & NGC 4571  &  & \coa{12}{36}{56.4} & \cod{+14}{13}{03}     &  SA(r)d	     & 11.73  &  342  & \amd{2}{4}  & A & 0.54	\\ 
1727 & NGC 4579  & M58 & \coa{12}{37}{43.6} & \cod{+11}{49}{05}  & SAB(rs)b	     & 10.29  & 1519  & \amd{1}{8}  & A & 0.83	\\ 
1730 & NGC 4580  &  & \coa{12}{37}{48.4} & \cod{+05}{22}{06}     & SAB(rs)a pec      & 12.49  & 1034  & \amd{7}{2}  & S & 1.03   \\ 
1943 & NGC 4639  &  & \coa{12}{42}{52.4} & \cod{+13}{15}{27}     & SAB(rs)bc	     & 11.85  & 1010  & \amd{3}{1}  & E & 0.25    \\ 
1987 & NGC 4654  &  & \coa{12}{43}{56.6} & \cod{+13}{07}{35}     & SAB(rs)cd	     & 10.75  & 1037  & \amd{3}{4}  & E & -0.29    \\ 
2058 & NGC 4689  &  & \coa{12}{47}{45.5} & \cod{+13}{45}{46}     & SA(rs)bc	     & 11.39  & 1616  & \amd{4}{5}  & E & 0.90    \\ 
\hline
\end{tabular}

\medskip
Notes on columns:
\ (1-3): Galaxy VCC, NGC-IC and Messier names.
\ (4-5): Right ascension and declination (J2000).  
\ (6): Morphological type, from RC3 (de Vaucouleurs et al. 1991).
\ (7): Total $B$-band magnitude, from RC3.
\ (8): Systemic velocity (in \kmps), from LEDA (Paturel et al. 1991).
\ (9): Projected distance to M87 (in degree).
\ (10): Revised region  of membership, as defined in Gavazzi et al. (1999): A = Cluster A, B = Cluster B, 
E = East cloud,  M = M cloud, N = North cloud, S = Southern extension.
\ (11): \hi\ deficiency parameter from Gavazzi et al. (2005). 
$^{a}$The value is uncertain for NGC~4438 due to uncertainties on its morphology and optical diameter.

\end{minipage}
\end{table*}

 The Virgo cluster of galaxies is the  nearest cluster from the Milky Way 
(D $\sim$ 16 Mpc) and is therefore very well appropriate to observe in details the galaxy properties. 
The Virgo Cluster Catalog (hereafter VCC, Binggeli, Sandage \& 
Tammann 1985; Binggeli, Popescu \& Tammann 1993) counts   $\sim 1400$ members 
with a dominating population of dwarf ellipticals.  
The cluster is not yet relaxed, as shown by the presence of irregular structures in either kinematical 
or X-ray data (Binggeli et al. 1987;  B{\" o}hringer et al. 1994; Shibata et al. 2001). The 3D 
structure of the cluster is very particular (see e.g. Gavazzi et al. 1999), articulated around a main substructure 
centered on Messier 87 (cluster A) and with several extensions.

 The selection criterium was to observe targets among the brightest spiral galaxies of the cluster. 
 This turns out to select objects having a total magnitude in the $B-$band 
 brighter than  $B^0_t = 12$. Other galaxies were observed because of their  
 presumably perturbed major-axis kinematics, as derived from long-slit spectroscopy  (Rubin et al. 1999), 
 their peculiar morphology (e.g. IC 3225) or their membership within a galaxy pair (e.g. IC 3583).
 A concern was to specifically observe galaxies in the inclination range between 25\degr\ and 80\degr\ 
 (with the exception of NGC 4438) in order to avoid  problems of disc opacity at optical 
 wavelength for high inclinations and of determination of the circular velocities for low inclinations.
 The Fabry-Perot catalog is composed of 30 galaxies which 
 locations are shown on Fig.~\ref{figposvf} and properties given in Tab.~\ref{targchar}.
 The morphological types of the targets  are distributed through 3 S0/a, 1 Sa, 3 Sab, 3 Sb, 7 Sbc, 4 Sc, 3 Scd, 
 3 Sd, 1 Sdm, 1 Sm and 1 Im galaxies.  The targets are mainly located in the cluster core where environmental
 effects are supposed to be more important, but also in the cluster extension towards M49. 
 Deep \ha\ imagery is available for all galaxies from this sample 
 (Koopmann, Kenney \& Young 2001; Boselli \& Gavazzi 2002; Boselli et al. 2002;  
  Gavazzi et al. 2002, 2003).
  
Optical kinematical data for many individual objects or larger samples of Virgo galaxies have 
been presented in the past using  long-slit spectroscopy of  
the \ha\ and \nii\ emission lines (e.g. Chincarini \& de Souza 1985; Distefano et al. 1990; Sperandio et al. 1995;
 Rubin et al. 1999; Yoshida et al. 2004).  Except for few individual galaxies 
(e.g. NGC 4254, Phookun, Vogel \& Mundy 1993;  NGC 4321, Arsenault, Roy \& Boulesteix 1990, 
Knapen et al. 1995, Canzian \& Allen 1997; 
NGC 4522, Vollmer et al. 2000), the mapping of the 2D optical kinematics for several Virgo galaxies has 
never been done before this study. The kinematics of the molecular and neutral atomic gas have 
nevertheless already been mapped at millimeter and centimeter wavelengths. 
The FP catalog has 10 galaxies in common with the Virgo CO Survey catalog of 
(NGC 4212, 4254, 4501, 4535, 4536, 4548, 4569, 4579, 4654 and 4689; Sofue et al. 2003a,b), 6 with the BIMA Survey Of Nearby Galaxies CO catalog 
(NGC 4321, 4450, 4535, 4548,  4569, 4579; Helfer et al. 2003), 15 with the \hi\ catalog (NGC 4237, 4254, 4321, 4438, 4450, 4501, 4535, 4548, 
4567, 4568, 4569, 4579, 4639, 4654, 4689; Guhathakurta et al. 1988; Cayatte et al. 1990, 1994) and other works like the one and two dimensional \hi\ data 
presented in Warmels (1988a,b,c). All of these will allow a comparison of the kinematics of the different gaseous components.

One has to note that not all galaxies of the sample lie at the same distance from us. 
For instance, accurate estimates based on the observations of Cepheid variable stars put NGC 4535 
at a distance of 16.0 $\pm$ 1.9 Mpc (Macri et al. 1999) and NGC 4639 at 25.5 $\pm$ 2.5 Mpc 
(Saha et al. 1997). The three-dimensional structure of the Virgo cluster has  often been studied (Yasuda, Fukugita \& Okamura 1997; 
Gavazzi et al. 1999; Fouqu\'e et al. 2001; Solanes et al. 2002) but is neglected here and  
for simplicity reasons it is assumed a distance of 16 Mpc for the cluster and all its galaxies. 
Therefore, 1\arcsec\ on the sky corresponds to $\sim$ 78 parsecs at this distance.


\section{Observations and data reduction}
\label{reduc}
\subsection{Data acquisition}
\label{acquistion}

\begin{table}
\centering
\caption{Telescope and Instrument configuration}
\label{teltel}
\begin{tabular}{lcccc}
\hline\hline
Telescope  & Detector  &  \fov & Detector & Pixel Size   \\
           &           &  (\arcmin)      &  sampling   &  (\arcsec)    \\
\hline
~A - OmM (1.60m)  & \fantomm  & 13.7    & $512\times512$  & 1.61  \\
B1 - OHP (1.93m) & \cigale    & 4.1     & $256\times256$  & 0.96  \\
B2 - OHP (1.93m) & \cigale    & 5.8     & $512\times512$  & 0.68  \\
C1 - CFHT (3.60m) & \fantomm  & 4.2     & $512\times512$  & 0.49  \\
C2 - CFHT (3.60m) & MOS/FP    & 4.9     & $512\times512$  & 0.57  \\
~D - ESO (3.60m)  & \fantomm  & 3.6     & $512\times512$  & 0.42  \\
\hline
\end{tabular}

\medskip
\end{table}
\begin{table*}
\begin{minipage}{145mm}
    \caption{Observational parameters}
    \label{targtel}
    \begin{tabular}{cccccccccccc}
\hline\hline
Galaxy & Set-up & Date &  $p$  &  $\rm \lambda_s$ & $\rm \lambda_f$ & $\rm FWHM_f$ & $\rm FSR$ & $nb_{\rm ch}$ & $t_{\rm exp}$ & $t$ & Seeing \\
    & (1)       & (2)  & (3)  & (4)             & (5)             & (6)          & (7)    &      (8) & (9)  & (10) & (11)  \\
\hline

 NGC 4189  & B2 & 07-03-03  & 793 & 6609.08 & 6612 & 11 &  380.71 & 24 & 5.0 & 120.0 & $\star$ \\
 NGC 4212  & B2 & 27-05-01  & 793 & 6561.01 & 6561 & 12 &  377.95 & 24 & 3.7 & 88.8  & $\star$     \\
 NGC 4237  & B2 & 10-03-03  & 793 & 6581.76 & 6582 & 11 &  379.14 & 24 & 5.2 & 124.8 & $\star$  \\
 NGC 4254  & A  & 14-02-05  & 765 & 6615.47 & 6621 & 18 &  395.03 & 48 & 5.0 & 240.0 & $\star$  \\
 NGC 4294  & C1 & 07-04-03  & 899 & 6570.64 & 6575 & 20 &  333.87 & 48 & 2.5 & 120.0 & $\star\star$	  \\
 NGC 4298  & D  & 06-04-02  & 793 & 6587.63 & 6589 & 12 &  379.48 & 24 & 2.3 & 55.2  & $\star\star$	 \\
  IC 3225  & D  & 04-04-02  & 609 & 6614.51 & 6617 & 11 &  496.15 & 40 & 2.3 & 92.0  & $\star\star$	 \\
 NGC 4321  & A  & 25-02-03  & 899 & 6597.17 & 6605 & 15 &  335.22 & 52 & 5.0 & 260.0 & $\star$ 	\\
 NGC 4351  & B2 & 27-04-03  & 793 & 6613.35 & 6614 & 11 &  380.96 & 24 & 4.5 & 108.0 & $\star$     \\
 NGC 4405  & B2 & 28-04-03  & 793 & 6601.02 & 6602 & 12 &  380.25 & 24 & 4.5 & 108.0 & $\star$        \\
 NGC 4438  & D  & 06-04-02  & 793 & 6564.33 & 6568 & 12 &  378.14 & 24 & 6.5 & 156.0 & $\star\star$	     \\
 NGC 4450  & D  & 07-04-02  & 793 & 6605.56 & 6607 & 12 &  380.51 & 24 & 2.5 &  60.0 & $\star\star$	     \\
 NGC 4457  & B2 & 10-03-03  & 793 & 6582.09 & 6582 & 11 &  379.16 & 24 & 5.2 & 124.8 & $\star$		\\
 NGC 4498  & B1 & 03-03-00  & 793 & 6595.77 & 6593 & 10 &  379.95 & 24 & 3.2 &  76.8 & $\star$		\\
 NGC 4501  & B1 & 10-03-00  & 793 & 6612.71 & 6614 & 11 &  380.93 & 24 & 4.8 & 115.2 & $\star$		 \\
 NGC 4519  & C1 & 04-04-03  & 899 & 6589.49 & 6595 & 18 &  334.83 & 48 & 2.0 &  96.0 & $\star\star$	   \\
 NGC 4532  & D  & 05-04-02  & 793 & 6606.82 & 6607 & 12 &  380.59 & 24 & 2.0 &  48.0 & $\star\star$	\\
 NGC 4535  & A  & 06-03-03  & 899 & 6605.71 & 6617 & 15 &  335.66 & 52 & 3.0 & 156.0 & $\star$ 	\\
 NGC 4536  & A  & 14-03-04  & 765 & 6602.36 & 6595 & 18 &  394.25 & 48 & 3.4 & 163.2 & $\star$ 	\\
 NGC 4548  & A  & 07-03-03  & 899 & 6573.42 & 6585 & 16 &  334.01 & 52 & 2.1 & 109.2 & $\star$		 \\
 NGC 4567  & B1 & 12-04-99  & 793 & 6612.56 & 6614 & 11 &  380.92 & 24 & 5.0 & 120.0 & $\star$		\\
 NGC 4568  & B1 & 12-04-99  & 793 & 6612.56 & 6614 & 11 &  380.92 & 24 & 5.0 & 120.0 & $\star$		\\
  IC 3583  & D  & 05-04-02  & 793 & 6587.32 & 6588 & 11 &  379.46 & 24 & 5.0 & 120.0 & $\star\star$   \\
 NGC 4569  & A  & 11-03-02  & 765 & 6557.64 & 6569 & 15 &  391.58 & 40 & 3.8 & 152.0 & $\star$		  \\
 NGC 4571  & B1 & 08-03-00  & 793 & 6570.27 & 6574 & 12 &  378.48 & 24 & 5.0 & 120.0 & $\star$ 	   \\
 NGC 4579  & D  & 04-04-02  & 609 & 6596.03 & 6598 & 10 &  494.76 & 40 & 2.3 &  92.0 & $\star\star$   \\
 NGC 4580  & B2 & 08-03-03  & 793 & 6585.42 & 6584 & 12 &  379.35 & 24 & 5.5 & 132.0 & $\star$ 	 \\
 NGC 4639  & D  & 05-04-02  & 793 & 6584.89 & 6588 & 11 &  379.32 & 24 & 3.3 &  79.2 & $\star\star$   \\
 NGC 4654  & C2 & 13-02-02  & 1162& 6585.48 & 6588 & 11 &  258.89 & 24 & 2.5 &  60.0 & $\star\star$	\\
 NGC 4689  & A  & 03-03-03  & 899 & 6598.16 & 6595 & 18 &  335.27 & 52 & 3.2 & 166.4 & $\star$ 	\\
\hline
\end{tabular}

\medskip
Notes on columns:
\ (1): Observing configuration : see Table~\ref{teltel} for details on the telescope and
instrument used during the observations.
\ (2): Date of observation (dd-mm-yy).
\ (3): Interference order (at the \ha\ wavelength).
\ (4): Scanning wavelength (\AA).
\ (5-6): Interference filter central wavelength and FWHM (in \AA, at the ambient temperature).
\ (7): Free Spectral range (at the scanning wavelength, in \kmps).
\ (8): Number of channels.
\ (9): Total exposure time per channel (in min).
\ (10): Total scanning exposure time (in min).
\ (11):  $\star\star$ is for a seeing $<$ \as{1}{5}, $\star$  for a seeing $\ga$ \as{2}{0} and $\la$ \as{4}{0}.
\end{minipage}
\end{table*}

The observations of the 30 galaxies were performed from 1999 to 2005 at
the ESO 3.6m (Chile), the 3.6m Canada-France-Hawaii (CFHT, Hawaii), the 1.93m Observatoire de Haute-Provence (OHP, France) and
the 1.6m Observatoire du mont M\'egantic (OmM, Qu\'ebec, Canada) telescopes. 
Each of our visitor instruments, \fantomm\ (at ESO, CFHT and OmM) and  \cigale\ (at OHP) is a package including 
a scanning Fabry-Perot interferometer, a series of 
interference filters and an Image Photon Counting  system  (IPCS) based on a technology of GaAs amplifier tube having a high sensitvity 
and a quantum efficiency of $\sim 28 \%$ (Gach et al. 2002; Hernandez et al. 2003). 
Our focal reducer (named \textit{Cigale} at ESO and OHP, \textit{MOS-FP} at CFHT, \textit{Panoramix}) 
is attached at the Cassegrain focus of the telescopes and is equipped with our visitor intrument. Table~\ref{teltel} and Table~\ref{targtel} give the observational set-up at each telescope
 and parameters of each galaxy (respectively). Among the 30 galaxies, NGC 4654 is the only one that has not been 
observed with an IPCS but with the CCD detector of the MOS/FP module at CFHT. 

To reconstruct a spectral profile in a spatial pixel during an exposure,  the spectral range (hereafter referred as $FSR$ for Free Spectral Range) 
of a FP interferometer is scanned through a number of channels ($nb_{\rm ch}$).
This number  must satisfy the Nyquist criteria, $nb_{\rm ch} \ga 2 \times \textit{F}$, where $F$ is the interferometer finesse. 
The finesse is related to the   resolving power $R$ of the interferometer by $R = p_s \times F$, where $p_s$ is the interference order 
at the scanning wavelength ($\lambda_{s}$). 
The spectral range at  $\lambda_{s}$ is given by $FSR = \frac{\lambda_{s}}{p_s}$ (in \AA) or 
$FSR = \frac{c}{p_s}$ (in \kmps), where $c$ is the speed of the light. 
 Columns 3,7 and 8 of Table~\ref{targtel} give the interference order (at \ha), 
the free spectral range (at $\lambda_{s}$) and the number of channels of the observations.
At the \ha\ wavelength, the finesses of the four used interferometers with interference orders of $p=$ 609, 765,
793, 899 and 1162 are between $\sim$10 and 23. This leads to resolving powers and spectral resolutions ranging between 
$\sim$ 7950 or 0.83 \AA\ ($p=$ 793 interferometer) and 
21000 or 0.31 \AA\ ($p=$ 899 interferometer). As a comparison, the resolution of long-slit spectroscopy observations (Rubin et al. 1999) is 
of the order of 1 \AA.

During an integration, each channel is scanned many times within a typical time of 15 seconds. When  
the sky transparency is not excellent prior to an exposure, the integration time is set to 10 seconds per channel 
in order to average more efficiently the effects of transparency changes. 
The important time for a FP observation is the total exposure time of a spectral sampling element,  
i.e. the total integration time of one channel ($t_{\rm exp}$, column 9 of Table~\ref{targtel}). 
The average overall integration time of the whole sample is 3.8 minutes per channel. 
The longest times are spent for galaxies observed at the 
smaller telescopes and for an object like NGC 4438,  
which has a very low surface brightness, diffuse and filamentary \ha\ morphology.    
To ensure an optimal wavelength calibration, a reference datacube of a Neon emission-line ($\lambda = 6598.95$ \AA) 
is obtained before and after the observation of a galaxy. 

\subsection{Data reduction}
\label{reduction}

The Fabry-Perot data were all reduced following a homogeneous procedure
which proceeds as follows: integration and wavelength calibration of the raw datacube, spectral
smoothing (hanning filtering) of the wavelength-calibrated datacube,
 night-sky emission-lines (hereafter NSEL, generally OH lines) subtraction, treatment of the pixels signal-to-noise ($S/N$) 
 and computation of the \ha\ integrated emission and velocity maps.
 The $\lambda$-calibration is done relatively to the above-mentionned Neon emission-line. 
Notice that no flux calibration of the datacubes is done.

We refer to other papers for a more detailed description of the wavelength calibration (e.g.  Laval et al. 1987; 
Amram  1991). The reduction package consists of \textit{IDL}\footnote{http://www.rsinc.com} routines that were adapted and improved 
from the \textit{adhocw}  software (Boulesteix 1993). The major improvements are for the sky subtraction and the signal-to-noise ratio ($S/N$) treatment 
procedures which are explained in detail and illustrated in Daigle et al. (2005b).
A brief summary of these major implementations for our data reduction follows. 
\begin{description}
\item[- \textbf{Night-sky emission-lines subtraction:}]
The often-used sky subtraction method of the \textit{adhocw} software consists in modelling a \emph{single NSEL average profile} which is
 extracted from \fov\ areas free from galaxy emission and removed from the observation. 
 Instead of this ``classical" NSEL subtraction method, the new procedure  constructs a \emph{NSEL datacube} which 
is fitted  to the whole field-of-view (\fov) and then subtracted from the observed datacube. 
Modelling the sky emission for each pixel instead of a unique pixel model reduces the NSEL residuals and thus allows a better
detection of galaxy emission in low-surface brightness regions (e.g. interarm, external regions...). 
Notice that if the sky-model fit cannot converge to an adequate solution, which can happen when there are two competing NSELs 
or when the galaxy emission covers a very large majority of the \fov, the
``classical"  method is used. This has been done e.g. for NGC 4438 and NGC 4579.
\item[- \textbf{Datacube signal-to-noise treatment:}]
An  adaptive spatial binning is used in order to increase the $S/N$ of the data, instead of the often-used gaussian smoothing. 
For that purpose \textit{IDL} routines were adapted to our Fabry-Perot data from existing routines 
developed by Cappellari \& Copin (2003) for the \sauron\ data (de Zeeuw et al. 2002). The 
method is a Voronoi tessellation and basically consists in 
spatially binning pixels with their neighbouring pixels until a bin with sufficient $S/N$ is obtained.  
 Consequently,
in regions of initial high $S/N$ (\hii\ regions in the spiral arms, in the disc,
in the galactic centre, ...), no spatial   binning   (or a minimum binning) is done and the angular
resolution is kept maximal in these regions, contrary to the gaussian smoothing method.
On another hand, because the binning occurs for pixels having a $S/N$ lower than the target $S/N$ value 
(low surface  brightness and diffuse gas in interam or outermost regions, ...), it allows to recover 
the spectral information over a larger part of a \fov\ than the gaussian smoothing method. 
The data presented here have a minimum $S/N$ of 5 over the field-of-view. An example of a binned velocity 
field is shown for the galaxy NGC 4254 in Fig.~\ref{vfbin} with spectra of bins (having different sizes) selected  
at some locations inside and outside the disc.
\begin{figure}
\begin{center}
  \includegraphics[width=\columnwidth]{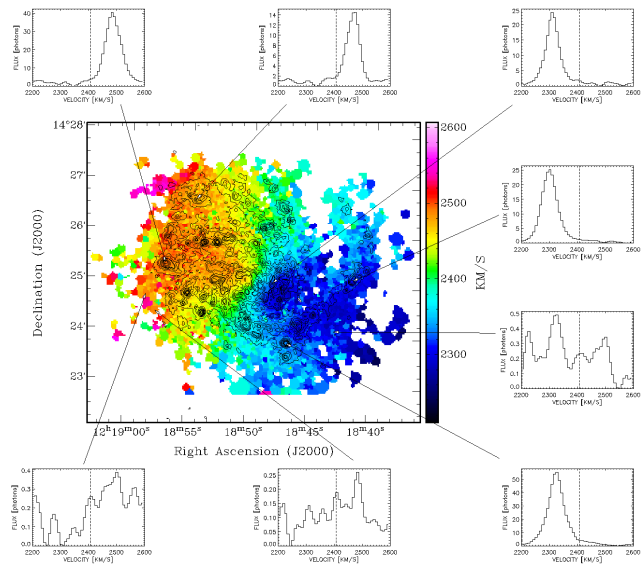}
\end{center}
\caption[]{\ha\ binned velocity field of NGC 4254 obtained by the Voronoi tessellation. 
The contours represent the \ha\ integrated emission of the galaxy. For each spectra, a dashed line 
corresponds to the systemic velocity of NGC 4254.}
\label{vfbin}
\end{figure}

\end{description}

For 3 galaxies (NGC 4405, NGC 4438 and NGC 4569),  a gaussian filtering of 3$\times$3 pixels FWHM (3$\times$3 and 6$\times$6 pixels 
for NGC 4438) was used instead of the Voronoi tessellation because  the solution of this new procedure  was not optimal for them. 
It can happen when the binning is done on a  NSEL residual (and so not on the \ha\ line of the galaxy) in extremely low surface brightness 
regions (e.g. filaments of NGC 4438, external arm of NGC 4569).  

The CCD data of NGC 4654 had two additional pre-processing procedures.  
Indeed, the CCD data require a cosmic ray hits subtraction and a photometric correction, procedures which are not 
yet implemented in our IDL routines.
Because each channel is scanned only one time during a CCD exposure  with an integration of several minutes per channel, 
significant sky transparency variations  can occur during an exposure and a photometric correction to the data is needed 
to take into account the sky flux changes.
 This problem does not exist with a photon counting system because each channel is quickly  scanned several times during an exposure
so that the transparency variations  are averaged from the beginning to the end of the overall integration.
For the CCD observation of NGC 4654, the cosmic ray hits were removed and a photometric correction was applied  
to the raw datacube as explained in Boulesteix (1993) before the wavelength calibration. 

The final steps of the data reduction consists in integrating the flux
under the \ha\ emission line, deriving its barycentre, dispersion and underlying continuum level 
following the method described in Daigle et al. (2005b).
We are finally left with an integrated \ha\ emission-line map and  a \ha\ velocity field. All radial velocities presented in this 
article are given in the heliocentric rest frame and are not absolute values. 
Notice that no systemic velocities are given in this article; 
 their values which do not present any particular interest for this study are generally in very good agreement 
 with  the referred ones (column 8 of Tab.~\ref{targchar}) within less than 15 \kmps.
 
Regions of galaxies that exhibit several emission lines in their profiles have their velocity 
fixed to the one of the highest intensity component (e.g. NGC 4438, Chemin et al. 2005). 
Mutiple spectral components are often found in interacting systems (Amram et al. 2004; Rampazzo et al. 2005).  
The FP  continuum map roughly traces the underlying stellar continuum of the \ha\ line.
It is not as deep or accurate as broad $R$-band images  
because it is integrated over $\sim 5-10$ angstroms. The FP continuum images have actually 
very little interest. For this reason, no continuum images are proposed in the article.
Their use only concerns the recovery of the data astrometry (see next section). 

\subsection{Data astrometry}
\label{wcs}

No reference to the World Coordinate System (WCS) have been  
obtained during the observations. It is important to know how a datacube is oriented in order to derive
the major-axis position angle (hereafter \pa) of a disc and the astrometric solution has to be found for the FP data. 
The \textbf{koords} task in the \textit{KARMA} package (Gooch 1996) is used to recover it, using a $R$-band map 
and a FP underlying continuum map as reference and target images respectively. The task  provides
 a reference pixel, its angular size on the sky plane and a possible rotation angle to align the \fov\ 
 with the WCS axes. 
 All the FP images presented in Appendix~\ref{velmap} have their vertical axis  aligned with the North.
 The original FP \fov\ are  very well oriented with the WCS. The orientation angle of the FP \fov\ vertical axis
 to the North is lower than $0.5$\degr. The worst misalignments reach $\sim 2\degr$ and are 
 for 9 galaxies observed at OmM and OHP. The values of position angles of  major axis given in Tab.~\ref{kinpar}  
 have been corrected of such strong misalignments. 
  
\begin{figure*}
\begin{center}
  \includegraphics[width=13.5cm]{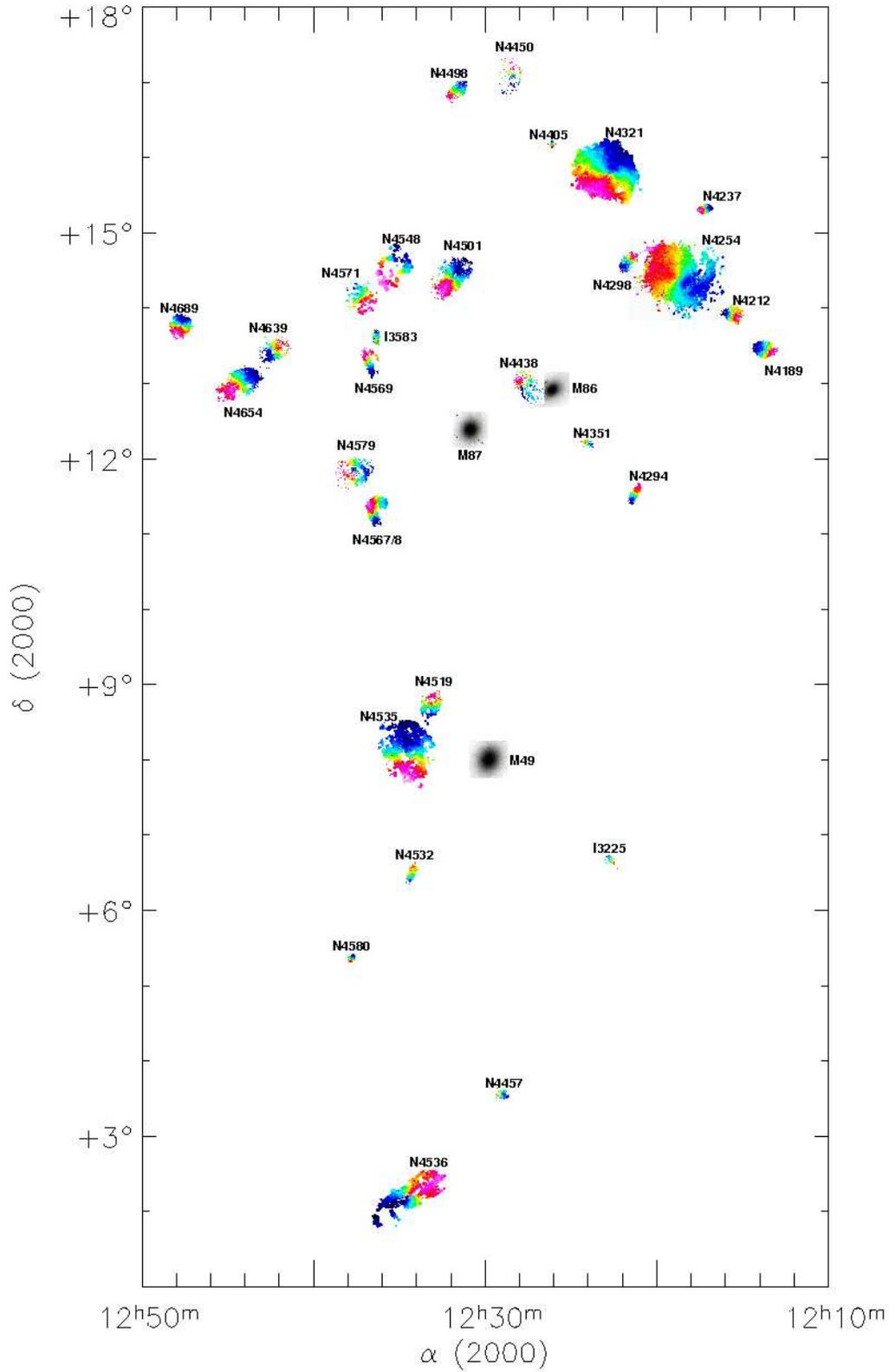}
\end{center}
\caption[]{A global view of the Virgo cluster galaxy velocity fields map. For clarity reasons, the angular scale of each galaxy has been enlarged 
by a factor of $\sim 9.5$ w.r.t. the real scale. The locations of the 3  ellipticals M49, M86 and M87 are displayed using  broad-band optical images. 
The colour scale is blue-darker shades (red-lighter shades) for the approaching (receding respectively) side of galaxies. See Appendix~\ref{velmap} for 
the velocity scale of each object.}
\label{figposvf}
\end{figure*}

\section{Results}
\label{results}
 \subsection{\ha\ emission line and kinematical maps}
 
  Figure~\ref{figposvf} gives a global representation of all 30 velocity fields from this sample positioned 
 at the location of galaxies in the Virgo cluster. 
Appendix~\ref{comments} presents a brief description of the observed kinematics for the galaxies of the sample. 
Appendix~\ref{velmap} presents their \ha\ emission line and kinematical maps. 

The catalog is arranged with increasing order of right ascension. 
For each galaxy, four images are displayed : a  $B$-band image (Top-Left), a $H$-band image (Top-Right), 
the FP \ha\ integrated map (Bottom-Left) and the corresponding \ha\ velocity field (Bottom-Right).  
The $B$- and $H$-band images are taken from the GOLDMine archive (Gavazzi et al. 2003).
For IC 3225, the $H$-band image is obtained from the 2MASS archive (Jarrett et al. 2000).
All four maps are WCS oriented and have the same field-of-view. 
The scales of the 4 images is chosen so that both the large scale stellar morphology and \ha\ distribution and kinematics 
are displayed with great detail, with the noticeable exception of NGC 4438.    
The position-velocity (PV) diagrams are also displayed, using the derived rotation curve projected onto the sky plane 
(solid line). The datacube slice has a width of 3-4 pixels. No PV diagrams are presented for NGC 4438 and NGC 4548 
for the reasons exposed in \S\ref{compphk}.
A constant value of \pa\  is adopted for NGC 4457, NGC 4501 and NGC 4579 despite the fact their kinematical \pa\ appeared to 
 vary with radius (see \S\ref{compphk}). 

Preliminary results for IC 3225, IC 3583, NGC 4298, NGC 4450, NGC 4532, NGC 4569, NGC 4579 and NGC 4639
have been presented in Chemin (2003) and detailed results for NGC 4438  in Chemin et al. (2005). 
This catalog presents the first optical kinematical data ever obtained for the galaxies IC 3225, IC 3583, NGC 4405 and NGC 4457. 
Forthcoming papers from this series will detail more quantitatively results
on the whole sample and on individual galaxies.

 The FP emission line maps are in agreement with the deep \ha\ images presented in Koopmann et al. (2001)  and in 
the GOLDMine database (Gavazzi et al. 2003).     
The discs have morphological perturbations such as nuclear spiral (NGC 4579), nuclear, inner and outer pseudo-ring 
(NGC 4321, NGC 4639), off-plane diffuse filaments, arc-like structure and giant \hii\ regions 
(NGC 4438, NGC 4532, NGC 4569), highly truncated \ha\ discs (NGC 4405, NGC 4569, NGC 4580), 
lopsidedness (e.g. IC 3225, NGC 4298, NGC 4654), $m=2$ modes such as a bar or 
two spiral arms, more complex spiral structure (NGC 4254, NGC 4535, NGC 4654). 
The signatures of kinematical perturbations are observed in the velocity fields in the form of  
streaming motions along spiral arms or pseudo-ring (e.g. NGC 4254, NGC 4294, NGC 4321, NGC 4639, NGC 4654, NGC 4689), 
steep velocity gradients along a bar or a nuclear structure (e.g. NGC 4519, NGC 4536, NGC 4579),  
Z-shape of the velocities due to the presence of a bar (e.g. NGC 4519, NGC 4532, NGC 4654), 
decoupled nuclear spiral (NGC 4579),  twist of the kinematical major axis (e.g. NGC 4501) or misalignment between
photometric and kinematical major axes (NGC 4254, NGC 4519), solid-body rise of the velocity 
in  truncated discs. Another (more subtle) perturbation is a kinematical lopsidedness 
that is only revealed by an in-depth  study of the velocity fields with titled-ring models and a harmonic decomposition
 (see Fig.~\ref{kincen} and section \ref{asymmetry}).  All these perturbations are probably responsible for 
 the observed irregularities of long-slit rotation curves (Rubin et al. 1999).

 \subsection{Kinematical parameters fitting}
An important step of the pre-analysis is to determine the location of the kinematical centre, the orientation parameters of a
 galactic disc (its inclination $i$ and angle of the major axis \pa) and the recession velocity $v_{\rm sys}$ of the 
disc from a kinematical map. Except for the inclination that is only a scale factor of the rotation curve and $v_{\rm sys}$, 
all the parameters are supposed known  \textit{before an observation} in long-slit spectroscopy studies.  
 The kinematical and photometric major axes are indeed supposed to be exactly the same and have a constant 
orientation across the galaxy, allowing the  
alignment of the slit with the  photometric major axis of a disc in order to derive its rotation curve. 
The dynamical centre is furthermore supposed to be at the location of the photometric one. 
It will be shown in the next section that these arguments sometimes do not hold and can lead to significant 
velocity (and thus mass) discrepancies with respect to the reality. 

There are several ways to derive a rotation curve $V_{\rm rot}$ from a velocity field.
Some 2D-kinematics optical studies fit by eye the parameters to the velocity field 
with efforts to lower the visual difference  
between the rotation curves of the  receding and approaching halves of a disc (e.g. Garrido et al. 2002). 
They also do not permit any variation of the parameters as a function of radius.    

In this work, it is decided  to use a more robust, rigourous and automatic fitting routine that fully uses the power of   
two dimensional maps and with  a minimum of human intervention. We use the \textbf{rotcur} task (Begeman 1987, 1989)
  of the \textit{gipsy} package  (van der Hulst et al. 1992). It considers purely axisymmetric motions by 
 fitting  the  formula $v_{\rm obs} = v_{\rm sys} + V_{\rm rot}.\sin(i).\cos(\theta) + V_{\rm rad}.\sin(i).\sin(\theta)$ 
to the observed velocity fields.   $V_{\rm rot}$ and $V_{\rm rad}$ are the circular and radial velocities in the 
galaxy plane. The terms  $\theta$, $i$, $v_{\rm obs}$ and $v_{\rm sys}$ refer to the azimuthal angle 
in the galaxy plane, the disc inclination,  the observed l-.o.-s. and systemic velocities (respectively).
The radial velocity is very frequently  not fitted in a velocity field 
(e.g.  Rogstad, Lockart \& Wright 1974; Bosma 1981; Begeman 1989;  
Carignan \& Beaulieu 1989; van der Hulst et al. 1993; 
de Blok, McGaugh \& van der Hulst 1996; Verheijen 2001) 
because it is considered very negligible with respect to the rotational term. 
Using a sample of 8 spiral galaxies observed in CO and \hi,   
Wong, Blitz \& Bosma (2004)  estimate that radial motions of 
inflow or outflow only reach an upper limit of 3-5\% of 
the circular velocity in only 3 velocity fields from their sample. 
It is therefore chosen here 
to neglect the $V_{\rm rad}$ term during the fittings.
    The parameters to be adjusted as a function of radius are the coordinates of the centre of mass, 
 $v_{\rm sys}$, $i$ and the position angle \pa\ of the major axis. This latter is the counterclockwise angle that the semi-major axis 
 of the receding half  does with the North axis. A good sampling of a ring implies its width to be set to a value which is at least 
 the average seeing  of the observation and $\ga$ 2 times the pixel size. It leads to ring width of 
$\ga$ 2\arcsec\ (for data collected at ESO and CFHT), 
 $\ga$ 3\arcsec\ (at OHP) and $\sim$ 5\arcsec\ (at M\'egantic). 
 Previously, the use of \textbf{rotcur} on \ha\ velocity fields was not very successful because of the poor spatial 
 coverage limited to the brightest \hii\ regions. But as can be seen in Appendix~\ref{velmap}, the spatial 
 coverage of these new data allows  \textbf{rotcur} to converge to accurate results, as with \hi\ data. 
  It is mainly due to the new generation of IPCS detector used and to the new adaptive binning technique that allows to get information of the 
 diffuse \ha\ component in interarm regions (see \S\ref{reduction}).  
 
 The sequence of the computation is similar to that usually used 
 for \hi\ velocity fields (e.g. Carignan \& Puche 1990;  Verheijen \& Sancisi 2001) and  proceeds as follows:
\begin{itemize}
\item  Average values of $v_{\rm sys}$ and of the dynamical center coordinates 
are found from their fitted radial profile.  
As an initial guess, the photometric parameters are used. 
The coordinates of the photometric centre were fitted to near-infrared (NIR) images obtained from the 
 2MASS and GOLDMine databases.  The average seeing of the NIR images is \as{1}{6} and the accuracy  
 on the photometric coordinates is less than 1\arcsec. The position of the photometric centre is 
 displayed with a cross in the figures of Appendix~\ref{velmap}. 
 The photometric position angles of the major axis and the inclination 
were taken in previous optical kinematical studies (Rubin et al. 1999), 
  in the LEDA database and the Uppsala General Catalog of Galaxies (Nilson 1973).  
Notice that in this first iteration the inclination is kept fixed at the photometric value. 
Keeping the value of \pa\ fixed or free  
does not influence the determination of $v_{\rm sys}$ and of the coordinates of the rotation centre. 
\item  Then, another least-square fitting is done with  
these new values of dynamical centre coordinates and $v_{\rm sys}$ as fixed parameters and  with  
$i$, \pa\ and the rotation velocity as free parameters. 
An average value of $i$ and \pa\ is computed from their fitted radial profile. 
\item  A last fitting is done by  having $i$ and \pa\ fixed by their new constant value, 
with the rotation velocity as the only free parameter. 
\end{itemize}
A residual velocity field is finally obtained by subtracting a model to the observation. 
This map is used to determine when an optimal  set of parameters is reached, i.e. when   the global 
average and the dispersion (r.m.s.)  of the residual velocities are minimum.

All points inside a half-sector of 30\degr\ (sometimes up to 50\degr\ for large inclinations) 
around the minor axis were rejected and a $|\rm \cos(\theta)|$ weight was applied to the l.-o.-s. velocities during the fittings (Begeman 1989). 
Both allow to minimize the contamination of the closest points to the minor axis where important projections effets occur 
and where the circular velocity term is fitted with difficulty ($\cos(\theta) \rightarrow 0$).
The average parameters are computed with some restrictions. First, one must avoid the inner regions 
where the velocities are  often affected by non-circular motions. This is noticeably the case 
for barred galaxies (Hernandez et al. 2005a). Their presence is clearly visible in the PV diagrams 
of Appendix~\ref{velmap}, where the  rotation curves rarely fit the  emission-line maxima in the central parts.      
 Then, one should not take into account rings which contain too few points. 
Results for such rings can give unrealistic or highly uncertain results when compared with other well-populated rings. 
This basically corresponds to rings having a number of points of less than $\sim$ 15\% that of the most 
populated ring, i.e. for the very inner and outer rings (see Fig.~\ref{exincl}, top panel).   
Finally, one should avoid  the regions where only one half of the disc is observed, or where the number of points 
in the fitting is highly dominated by that of only one side of a disc. 

The two ``siamese" galaxies NGC 4567 and NGC 4568 obviously superpose along the line-of-sight. The fittings should be partly affected  
by pixels of overlap in their receding halves. The method to decide which of the pixels belong to NGC 4567 or NGC 4568 was to fit models
 until the residual velocities are minimum in the region of overlap. As a result, it appears that   the
 region of overlap has minimum residual velocities when they are attributed to the NGC 4568 velocity field. This is 
 fully consistent with the photometry where NGC 4568 seems  to be projected at the foreground of the pair. 
 The   parameters of both galaxies are thus accurately determined.          

\begin{table}
\begin{minipage}{82mm}
    \caption{Virgo galaxy targets parameters}
    \label{kinpar}
    \begin{tabular}{crrrrcc} 
\hline\hline
Galaxy &  $P.A._{\rm p}$  & $i_{\rm p}$ &   $P.A._{\rm k}$  & \multicolumn{1}{c}{$i_{\rm k}$} & \multicolumn{2}{c}{Offset}  \\
  & (1)    & (2) & (3) & \multicolumn{1}{c}{(4)}   & (5) & (6) \\
\hline
 NGC 4189 & 265 & 45  & $ 251 \pm 6 $ &  $(31 \pm 6)$     &4.3 &335 \\
 NGC 4212 & 255 & 53  & $ 259 \pm 2 $  & $ 49 \pm 6 $     &1.3 &101  \\
 NGC 4237 & 108 & 51  & $ 106 \pm 2 $  & $ 53 \pm 3 $     &2.0 &156 \\
 NGC 4254 & 45  & 28   & $ 69 \pm 3 $  & $ 31 \pm 6 $     &5.2 &406 \\
 NGC 4294 & 335 & 71  & $ 335 \pm 3 $  & $ 68 \pm 6 $     &4.1 &320 \\
 NGC 4298 & 320 & 58  & $ 316 \pm 3 $  & $ 58 \pm 3 $     &1.6 &125 \\
  IC 3225 & 213 & 75   & $ 215 \pm 4 $ & $(70 \pm 12)$    &2.7 &211 \\
 NGC 4321 & 153 & 32  & $ 151 \pm 3 $  & $ 38 \pm 7 $     &6.6 &515  \\
 NGC 4351 & 80  & 49   & $ 73 \pm 5 $  &   $(39 \pm 54)$  &2.6 &203  \\
 NGC 4405 & 200 & 54   & $ 199 \pm 3 $  & $ 57\pm 6 $     &2.2 &172  \\
 NGC 4450 & 355 & 43  & $ 351 \pm 7 $  &  $(49 \pm 17)$   &3.6 &281 \\
 NGC 4457 & 82  & 34   &  $88 \pm 22 $ &  $(27 \pm 19)$   &6.1 &476 \\
 NGC 4498 & 133 & 59  & $  136 \pm 4 $  & $ 63 \pm 4 $    &7.7 &601   \\
 NGC 4501 & 140 & 59  & $ 141 \pm 2 $  & $ 58 \pm 6 $     &4.3 &335 \\
 NGC 4519 & 325 & 40  & $ 357 \pm 4 $  & $ 40 \pm 5 $     &7.3 &569  \\
 NGC 4532 & 340 & 64  & $ 355 \pm 6 $  &  $(63 \pm 14)$   &10.3&803    \\
 NGC 4535 & 180 & 46    & $ 180 \pm 2 $  & $ 38 \pm 10 $  &2.2 &172    \\
 NGC 4536 & 310 & 67  & $  301 \pm 2 $  & $ 68 \pm 3 $    &9.6 &749  \\
 NGC 4567 & 85 & 49  & $ 90 \pm 1 $  & $ 53 \pm 4 $       &2.7 &211   \\
 NGC 4568 & 23 & 67  & $ 27 \pm 3 $  & $ 67 \pm 3 $       &2.2 &172   \\
  IC 3583 & 180 & 72    & $ 178 \pm 5 $  &  $(77 \pm 11)$ &4.4 &343   \\
 NGC 4569 & 23.5 & 65 & $ 23 \pm 4 $  & $ 67 \pm 5 $      &4.0 &312    \\
 NGC 4571 & 220 & 28   & $ 220 \pm 4 $  &  $(37 \pm 17)$  &3.2 &252  \\
 NGC 4579 & 95 & 38  & $  91 \pm 3 $  & $ 44 \pm 8 $      &1.8 &140  \\
 NGC 4580 & 165 & 40  & $ 157 \pm 3 $  & $ 46 \pm 10 $    &0.9 & 70   \\
 NGC 4639 & 304 & 49  & $ 311 \pm 2 $  & $ 48 \pm 6 $     &2.4 &187   \\
 NGC 4654 & 128 & 57  & $ 125 \pm 2 $  & $ 62 \pm 7 $     &2.4 &187   \\
 NGC 4689 & 163 & 36  & $ 165 \pm 2 $  & $(36 \pm 14)$    &2.1 &164  	\\
\hline
\end{tabular}

\medskip
Notes on columns: \ (1-2): Photometric position angle $P.A._{\rm p}$ and inclination $i_{\rm p}$, (in degree). Taken from Rubin et al. (1999), LEDA, UGC. 
\ (3-4): Kinematical position angle $P.A._{\rm k}$ and inclination $i_{\rm k}$, (in degree). Numbers into 
parentheses  indicate that the kinematical inclination has not been  chosen to obtain the rotation curve but the photometric value.
\ (5-6): Sky projected offset between  the photometric and kinematical centres (in arcsec and parsec respectively).
\end{minipage}
\end{table}

\subsection{Comparison with photometric parameters}
\label{compphk}
\begin{figure}
\begin{center}
  \includegraphics[width=4cm]{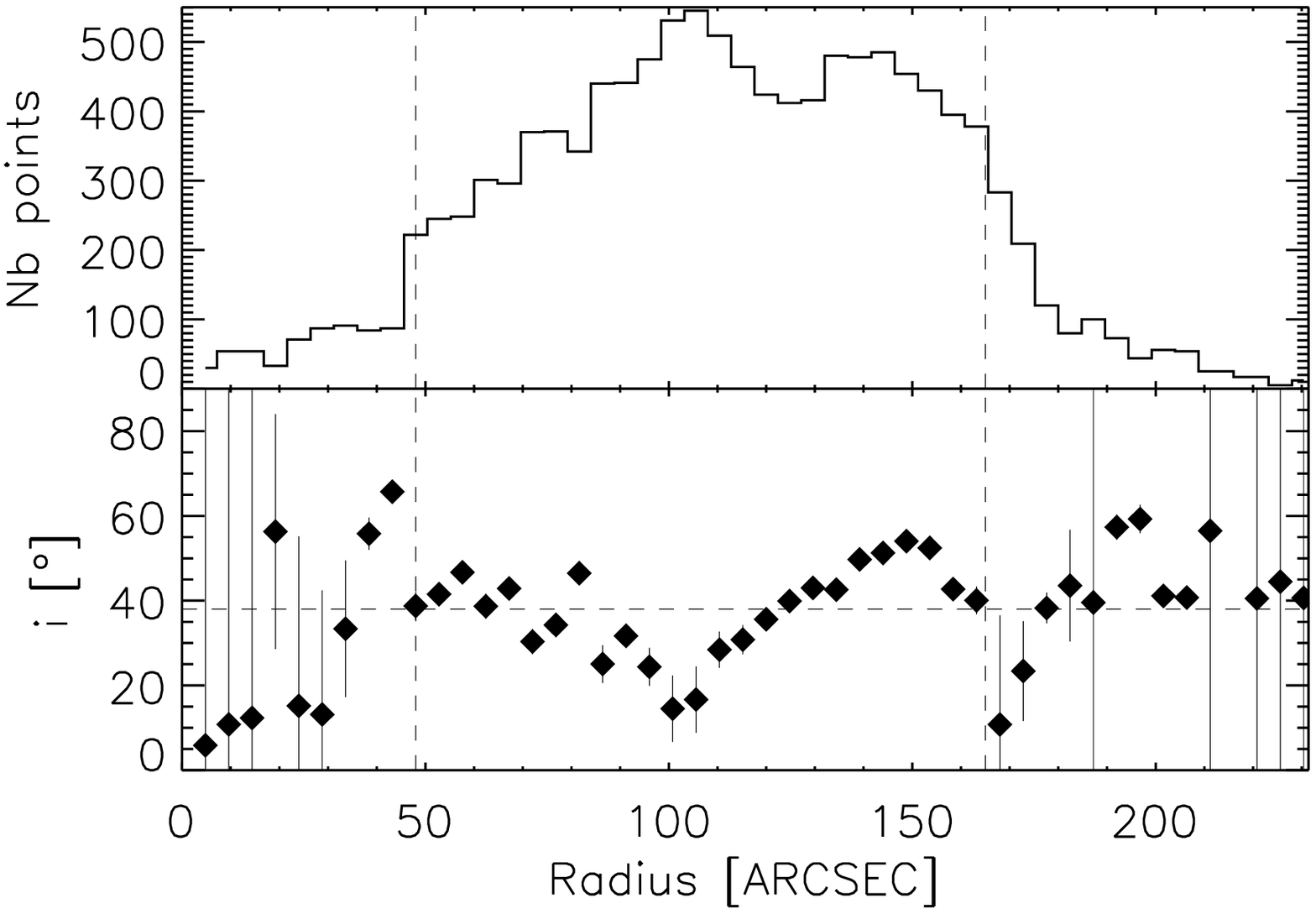}\includegraphics[width=4cm]{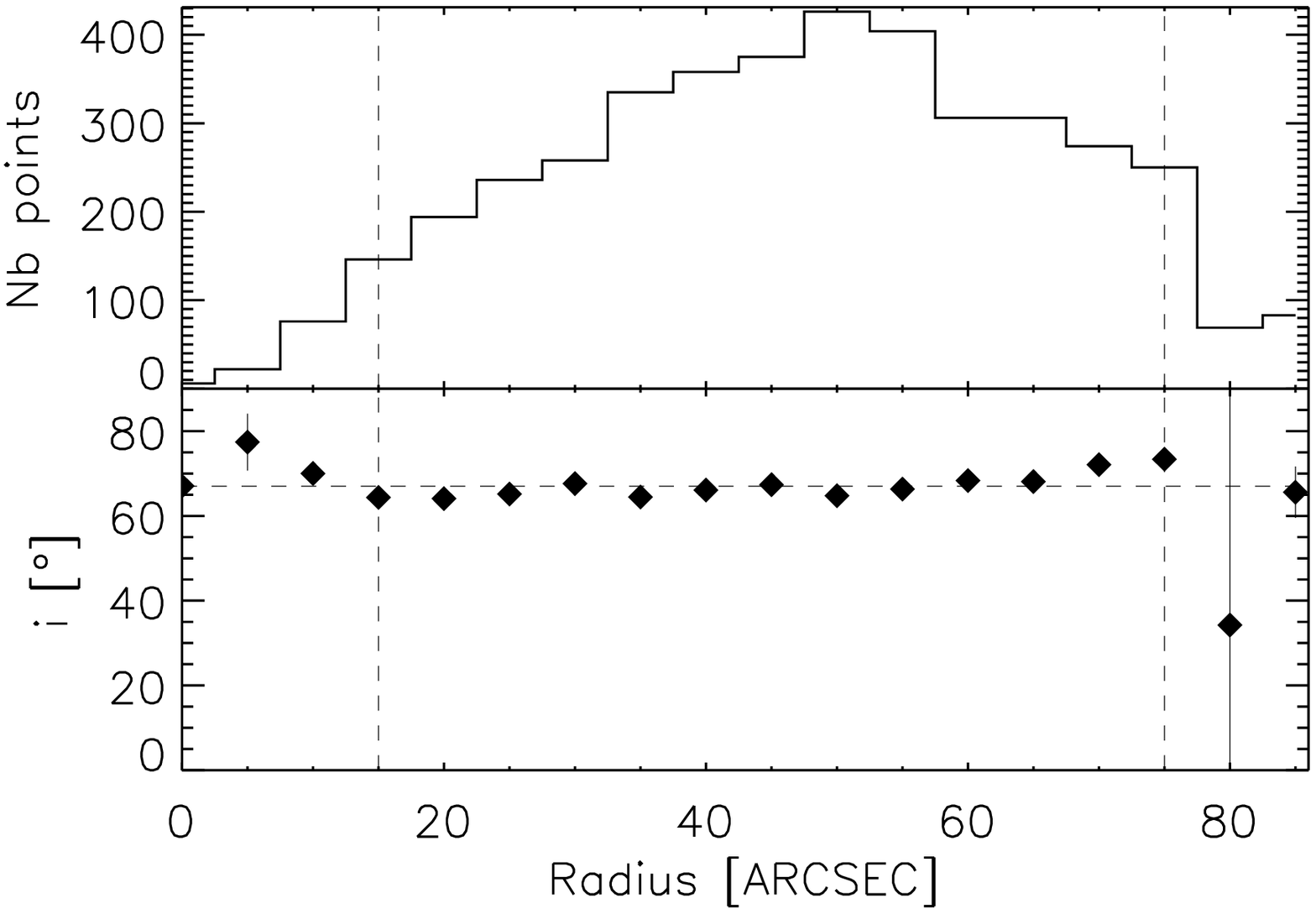}
  \includegraphics[width=4cm]{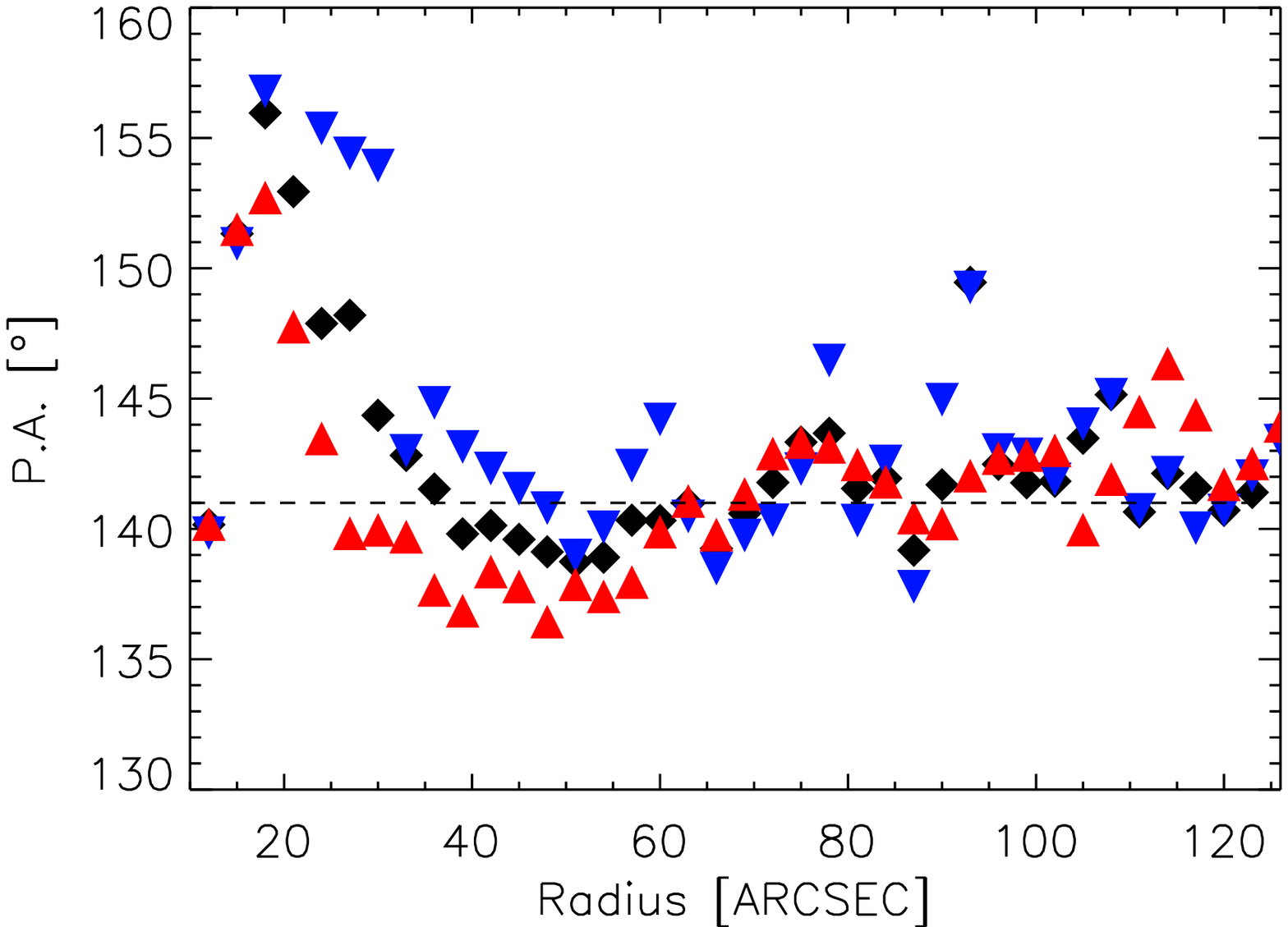}\includegraphics[width=4.0cm]{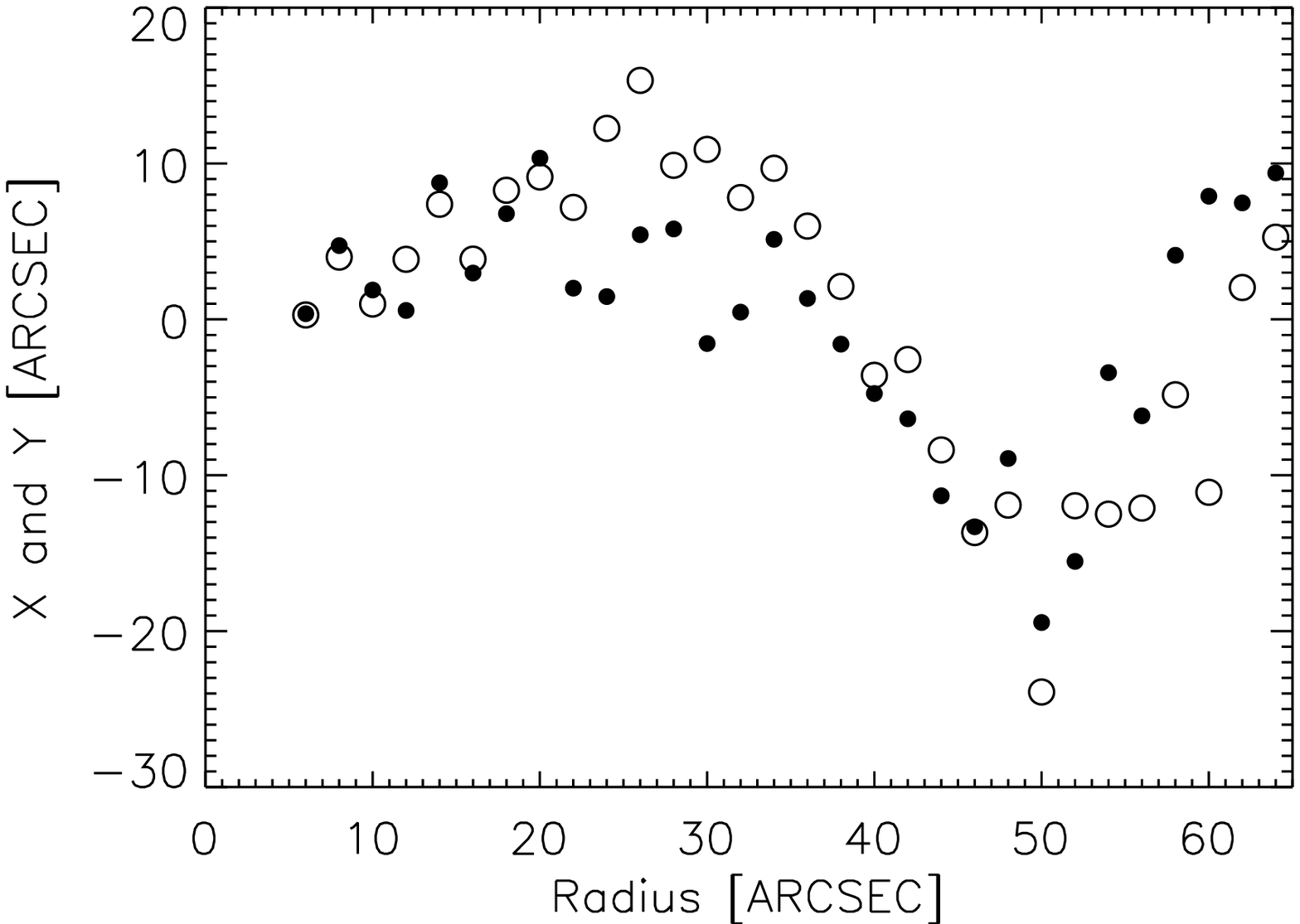}
\end{center}
\caption[]{(Top) Radial profiles of kinematical inclination for NGC 4535 (Left) and NGC 4568 (Right). A dashed horizontal 
line represents the mean inclination and two dashed vertical lines delineate the radial 
 range within which the mean value is computed. The top panels display the number of points of each annulus. 
 (Bottom-Left) Radial profile of kinematical position angle for NGC 4501. Blue (red) triangles 
are for the approaching (respectively receding) half of the disc and filled  diamonds for both sides 
fitted simultaneously.  (Bottom-right) Radial profile of the X- and Y-coordinates (open and full circles) 
of the kinematical centre for NGC 4298.}
\label{exincl}
\end{figure}

\begin{figure}
\begin{center}
  \includegraphics[width=4cm]{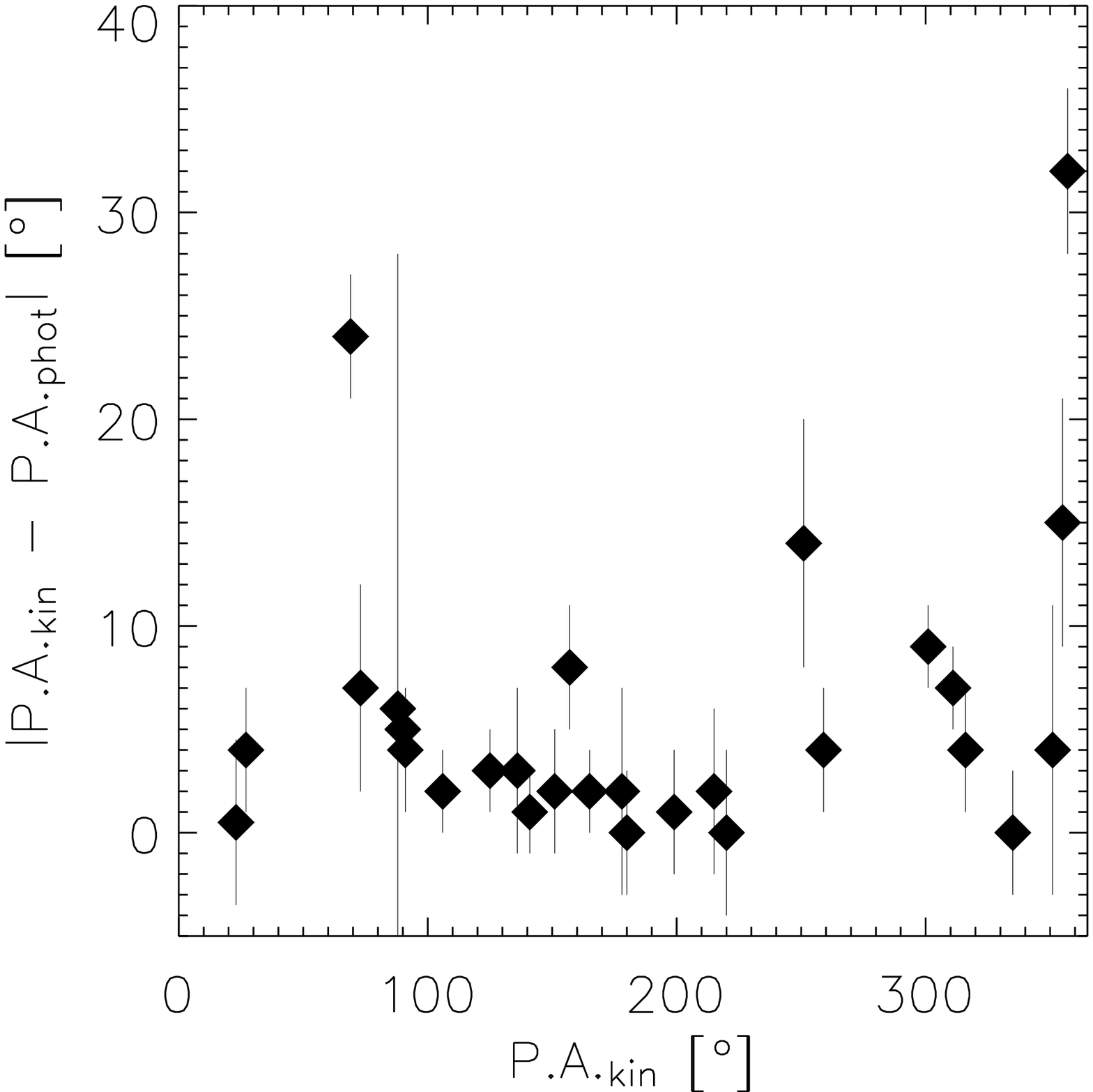}\includegraphics[width=4cm]{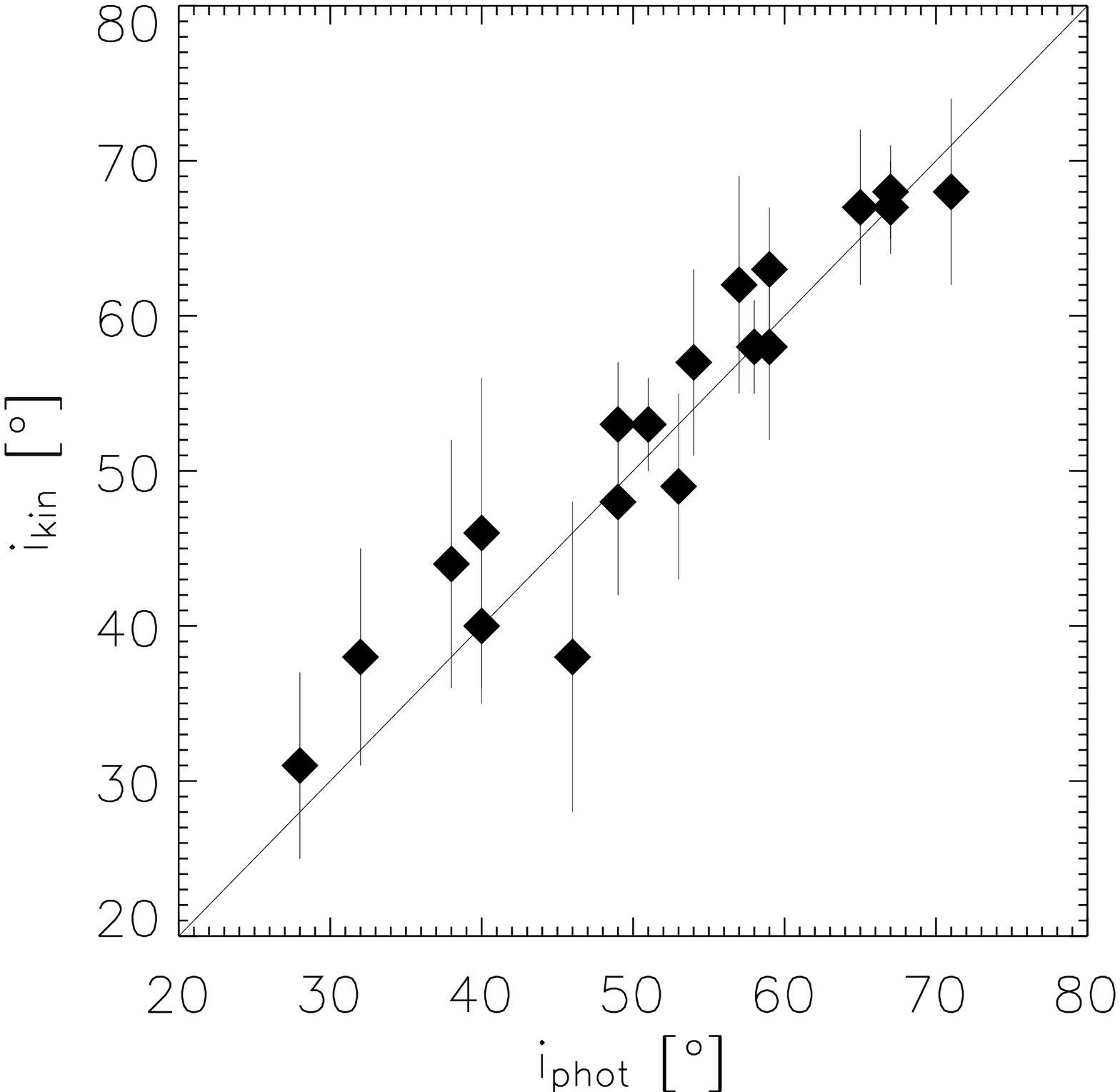}
\end{center}
\caption[]{Comparison between photometric and kinematical orientation parameters: absolute value 
of relative position angles versus the kinematical position angle (Left). Inclination (Right).}
\label{figinclpa}
\end{figure}

\begin{figure} 
\begin{center}
\includegraphics[width=3.3cm]{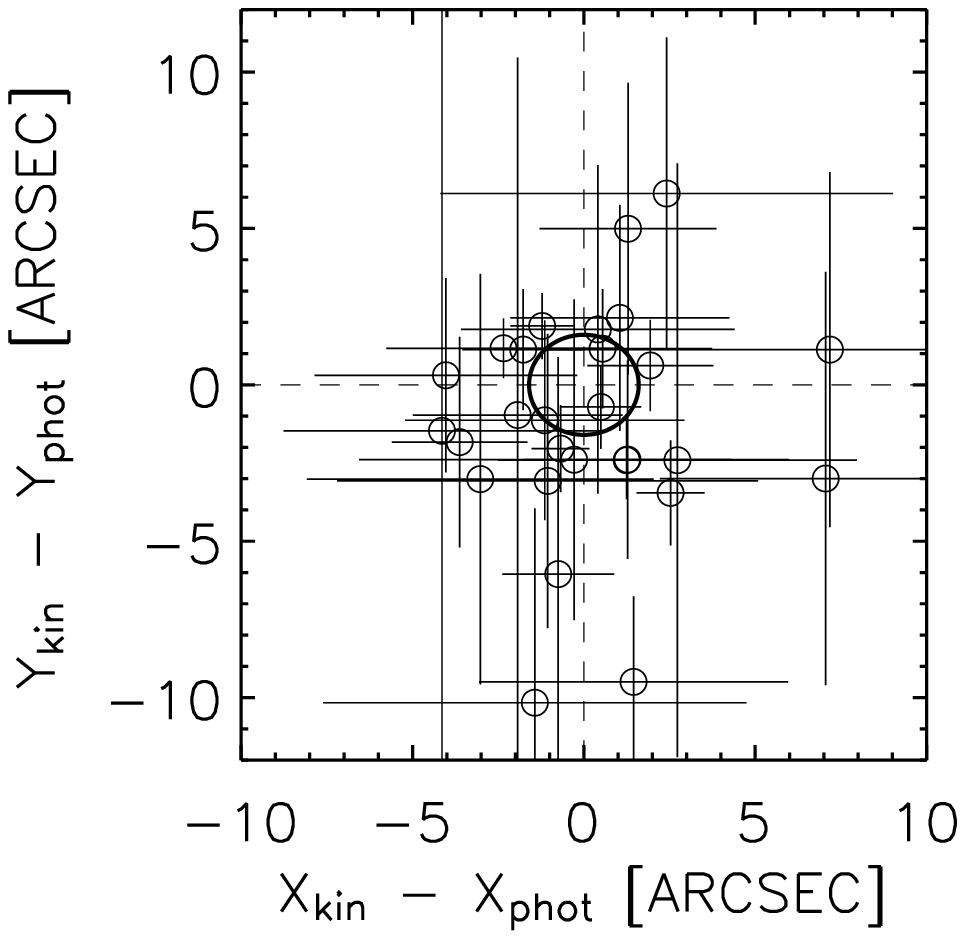}\includegraphics[width=4.9cm]{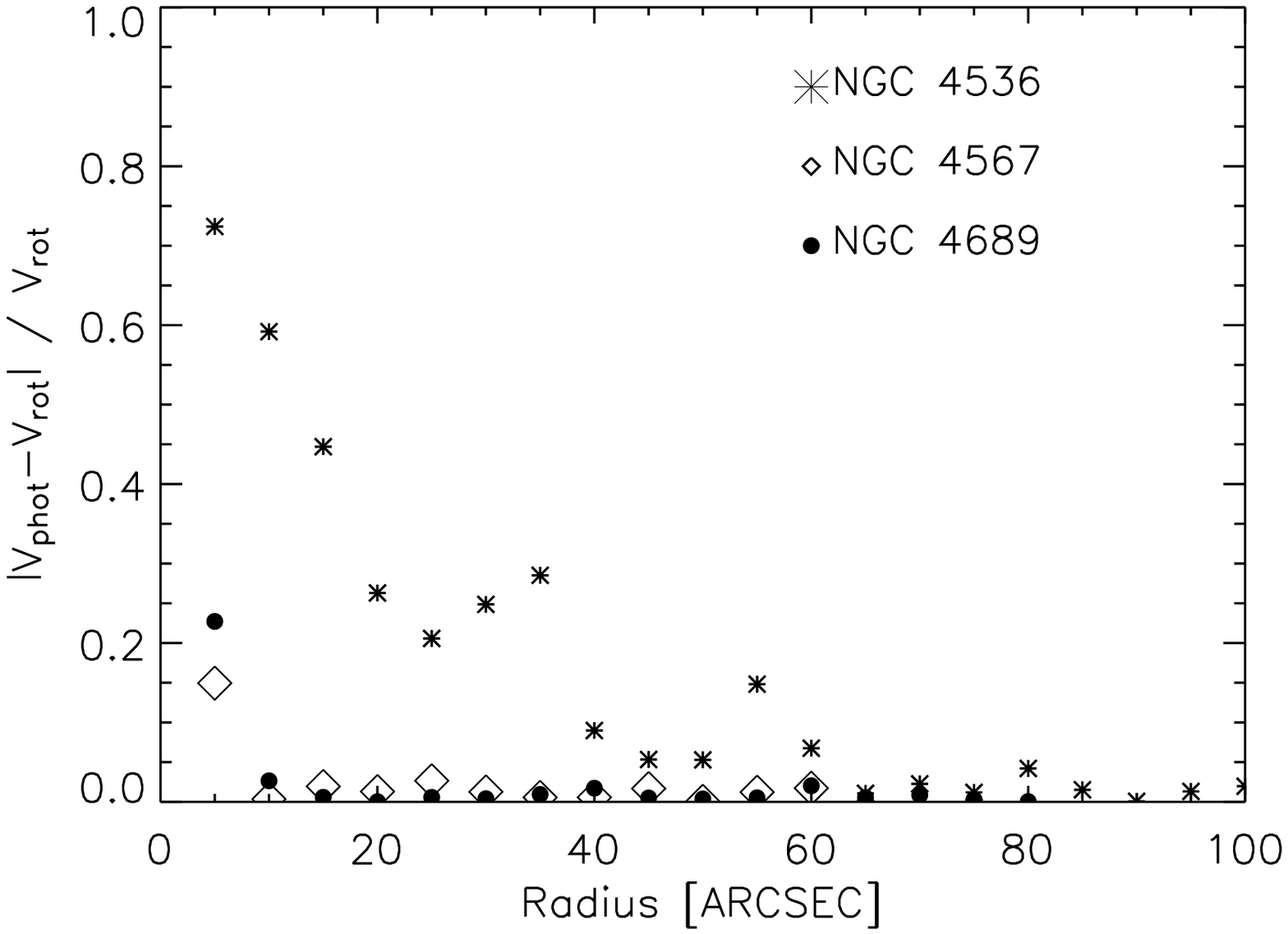}
\end{center}
\caption[]{(Left) Comparison between the kinematical and photometric coordinates. The horizontal and vertical bars are the r.m.s. of the kinematical
coordinates (resp.) and the circle centered on [0,0] is the mean seeing of the NIR plates, i.e. the error on the position of the
photometric centre. (Right) Comparison  between the rotation curves $V_{\rm rot}$ and $V_{\rm phot}$  for NGC 4567, NGC 4536 and NGC 4689. 
These curves are respectively obtained by fixing the rotation centre at the kinematical and photometric centres.}
\label{kincen}
\end{figure}

 Figure~\ref{exincl} illustrates examples of results from the tilted-ring model fittings to the velocity fields.
Table~\ref{kinpar} gives the kinematical orientation parameters. 
The error-bar given in Tab.~\ref{kinpar} is the r.m.s. (1$\sigma$-dispersion) of the radial profiles of the inclination or the position angle. 
Figure~\ref{figinclpa} shows a comparison between the photometric and kinematical 
inclinations and position angles of the major axis. A similar comparison is made for the position of the photometric and kinematical 
centres in Fig.~\ref{kincen}. 
No fitting has been done for the extremely perturbed galaxy NGC 4438 (as explained 
in Chemin et al. 2005) and for NGC 4548 (lack of independent points because of too large bins). No position-velocity diagrams 
are shown for these two galaxies.

\textbf{\emph{Position angle of the major axis.-}}
The position angle of the major axis is generally the best determined parameter, 
with a typical dispersion of 3\degr. 
The largest dispersions are for very perturbed galaxies (IC 3583, NGC 4532) or objects having 
a peculiar kinematics (NGC 4450).
 A comparison of the kinematical \pa\  with the photometric one is 
 done in Figure~\ref{figinclpa} and shows that there is a general  good agreement between 
 the two angles.  A few galaxies have a misalignment of more than 10\degr\ with the photometry
 (e.g. NGC 4189, NGC 4519). An analysis of their residual velocity maps  
 confirm that the  \pa\ derived from their velocity field is the angle 
 that  describes better the gas motions in the disc than when using the photometric \pa\ (see \S\ref{compkinphot}).   
  
 For differences lower than 10\degr\ (23 galaxies among 28 when removing NGC 4457 which has 
 a strong kinematical twist), the mean absolute difference between the two angles 
 is 3\degr, which is exactly the typical error-bar on the kinematical \pa. For those 
 galaxies, the velocity curves derived from long-slit data should give a good representation of the 
 rotational pattern, providing that the position of the photometric centre is also in agreement 
 with the dynamical one.  
 
 A few velocity fields exhibit significant kinematical twists (NGC 4457, NGC 4501 and NGC 4579). 
 Kinematical twists are often associated with a warp outside the optical disc 
 and thus with a change of inclination of the \hi\ disc (e.g. Sicotte \& Carignan 1997). 
 For NGC 4501 and NGC 4579, the twists occur in the galaxy centres. The NGC 4579 twist 
 is clearly linked to  its nuclear spiral (see Appendix~\ref{comments})  
 which orientation differs by $\sim 90\degr$ from that of the large-scale spiral arms. 
 An example  is shown in Fig.~\ref{exincl} (bottom) with NGC 4501. Such a twist 
 was also observed in CO data (Wong, Blitz \& Bosma 2004) and was attributed to an unkown, hidden bar. 
 Its presence remains however to be confirmed by photometry.
 For those galaxies, a mean value is derived within a  
 radial range where the angle variations are the smallest. It is this mean value which is given 
 in Tab.~\ref{kinpar} and used in the making of their PV-diagram.
  Notice that for NGC 4457, the given  \pa\ dispersion remains very large because the \pa\ 
 continuoulsy decreases with radius. For 
 the two remaining galaxies, the \pa\ estimates outside that central region of variation are 
 in good agreement with the photometric values. 
  
\textbf{\emph{Inclination.-}} 
The kinematical inclination is generally higher than the photometric value but   
 remains in agreement with the photometry within the errors. 
A large dispersion for the inclination illustrates the difficulty to accurately determine it from 
  high resolution velocity fields with a tilted-ring model, mainly due to the coupling   
of the circular velocity and the inclination  during the least-square fitting  
($V_{\rm rot}.\sin(i)$ term of the line-of-sight velocity). 
The largest dispersions are mainly for inclinations lower than $\sim 50\degr$, 
as already shown in Begeman (1989) with \hi\ data.  Moreover, as for the position angle, the 
inclination of mildly perturbed discs are less accurate than those having a more regular morphology. 
Figure~\ref{exincl}  shows examples of inclination fittings for the galaxies NGC 4535 ($\sigma = 10\degr$)
and NGC 4568 ($\sigma = 3\degr$). 
The kinematical inclinations for IC 3225, IC 3583, NGC 4189, NGC 4351, 
NGC 4450, NGC 4457, NGC 4532, NGC 4571 and NGC 4689 are very uncertain ($\sigma > 10\degr$) or very 
inconsistent with the photometric value (see Tab.~\ref{kinpar}). 
This is not surprising since all these galaxies are either morphologically perturbed discs 
or have a presumably low photometric inclination. The photometric value was thus used for the 
extraction of their rotation curve and the modelling of their velocity field.

It is worth noting that $m=2$ perturbations of the potential of a disc have been shown to also 
affect the results of inclination and \pa\ fittings  in  tilted-ring models    
(Schoenmakers, Franx \& de Zeeuw 1997; Schoenmakers 1999; Fridman et al. 2005). The parameter 
is observed to oscillate around its mean physical value and it results to an increase of  
the errors on it, as probably observed in many galaxies of this sample. 
Notice finally that  \hi\  velocity maps with lower resolution than the \ha\ observations logically give  
lower dispersion of the inclination  (e.g. Guhathakurta et al. 1988; Wong et al.  2004), 
which  parameters compare well with our FP estimates. 

 \emph{Position of the rotation centre.-} 
 An example of a radial profile of the kinematical coordinates  
is displayed in Figure~\ref{exincl} for the galaxy NGC 4298. 
The coordinates show  a large variation as a function of  radius.  
This implies a large dispersion  around the mean coordinates.   
 In view of this quite large variation of the coordinates, getting the 
  most accurate rotation  curve with an optical  velocity field should perhaps 
require to leave the centre vary as a function of radius. This is  indeed shown to give the most symmetric rotation curves
  (Beauvais \& Bothun 2001). The impact of the oscillations of the coordinates of the dynamical centre on a
 rotation curve is nevertheless not the purpose of this article. It will be studied in a forthcoming paper.
 As usually done, it has been  decided here to keep the centre fixed and at a constant position. 
 Keeping it  constant is precisely what allows to further detect  a kinematical lopsidedness in the harmonic analysis that could reflect 
 an environmental effect on a galaxy (see \S\ref{asymmetry}). \\
 Figure~\ref{kincen} shows the difference between the coordinates of the photometric and kinematical 
centres (see also in Tab.~\ref{kinpar} and the images of Appendix~\ref{velmap}).
The [0,0] location corresponds to that of the  photometric centre and a circle to its uncertainty. 
Its radius is thus given by the seeing of the NIR plates (\as{1}{6}). 
Though there are very few points inside this circle, the positions of the two centres seem 
consistent with each other due to the large scatter often observed around the kinematical values (as seen with NGC 4298). 
  The largest departures are for  perturbed galaxies (NGC 4457, NGC 4498, NGC 4519, NGC 4532)  
  but also for large, barred galaxies (NGC 4321, NGC 4536). 
  For all these objects, it is probably more difficult to determine the position of the photometric centre than for other 
  more regular galaxies.  Also, the extinction by dust near a galaxy nucleus could play a role in the determination of the 
  photometric centre, making it harder to find though NIR images are used here for that purpose.\\
Consequences of centering effects on a rotation curve have been discussed in Beauvais \& Bothun (2001). 
We qualitatively find the same result as them that when choosing the photometric centre instead of the kinematical centre, 
the velocities of the very innermost annuli of a rotation curve can change drastically. 
We illustrate this point in Figure~\ref{kincen} (right panel) with the galaxies NGC 4536, NGC 4567 and NGC 4689.
These objects were chosen because the departure between the centres exceeds the NIR plates seeing of \as{1}{6}.
The expression $\frac{|V_{\rm phot}-V_{\rm rot}|}{V_{\rm rot}}$ is displayed, where
 $V_{\rm phot}$ is the rotation curve derived by fixing the centre at the photometric position (using the kinematical inclination and $P.A.$). 
A systemic velocity corresponding to this centre was fitted and a corresponding 
residual velocity field was built. The r.m.s. of this residual map is generally by very few \kmps\ greater 
than that of the residual map built using $V_{\rm rot}$ and the kinematical centre.  
The optical centre is thus not adequately the centre of rotation, though  not being necessarily so far from it within 
the estimated error-bars.\\
 The graph of Figure~\ref{kincen}  shows that the difference between the two curves always decreases as a function of radius.
The curves only differ by $\sim$ 2\% in the outer regions. The  flat part of a rotation curve 
is therefore not affected by a minor change of centre.  In the innermost regions of barred galaxies,  differences of up to 70-80\%   are observed. 
 This difference is not significant  because the rotation velocity is highly uncertain, 
 whatever the choice of a centre. This is very likely due to important non-circular motions of the gas kinematics in the central parts 
 of barred galaxies. For non-barred late-type spirals, the velocity of the innermost annuli  differ by $\sim$ 5 to 25\%. 
 This value has to be compared with the formal error-bar  measured by \textbf{rotcur} at these radii, which is 
 less than 5\% of $V_{\rm rot}$. Therefore, contrary to the barred objects, the difference 
 is critical because the velocities at these radii are essential in the determination of 
  the disc mass-to-light ratio in mass models. The choice of the centre is important for this type of galaxies.  
\section{A short analysis}
\label{analysis}
This section briefly describes some goals that will be achieved with the high-resolution 3D optical data 
in future articles from this series.

\subsection{Comparison with other \ha\ works}
\label{compkinphot}
\begin{figure}
\begin{center}
  \includegraphics[width=\columnwidth]{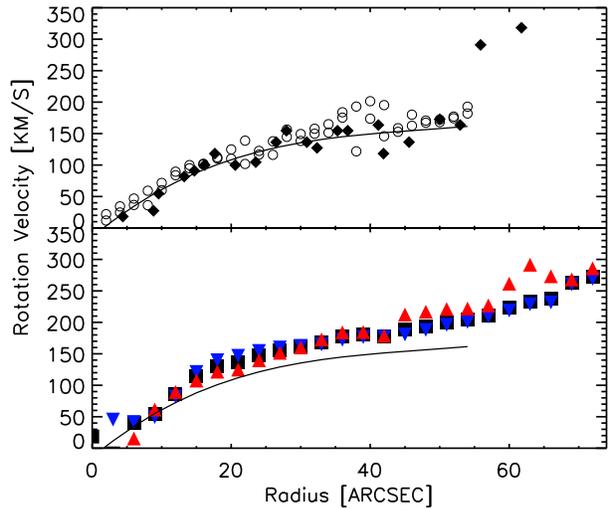}
\end{center}
\caption[]{Comparison between the rotation curves of NGC 4189. (Top panel)   
Long-slit rotation curve (Rubin et al. 1999, filled squares), smooth model of long-slit data (solid line) and rotation curve from  
the FP velocity field extracted along the photometric axis of 265\degr\ in a $\sim 2\arcsec$ width pseudo-slit (open circles).  
(Bottom panel) Rotation curve  of NGC 4189 derived from the 
tilted-ring model fitting to the velocity field (filled squares and triangles) for a value 
position angle of the major axis of 251\degr.  Blue (red) triangles 
are for the approaching (respectively receding) half of the disc. The solid line is the smooth model of long-slit data.}
\label{vrotn4189}
\end{figure}

\begin{figure*}
\centering
   \includegraphics[height=5cm]{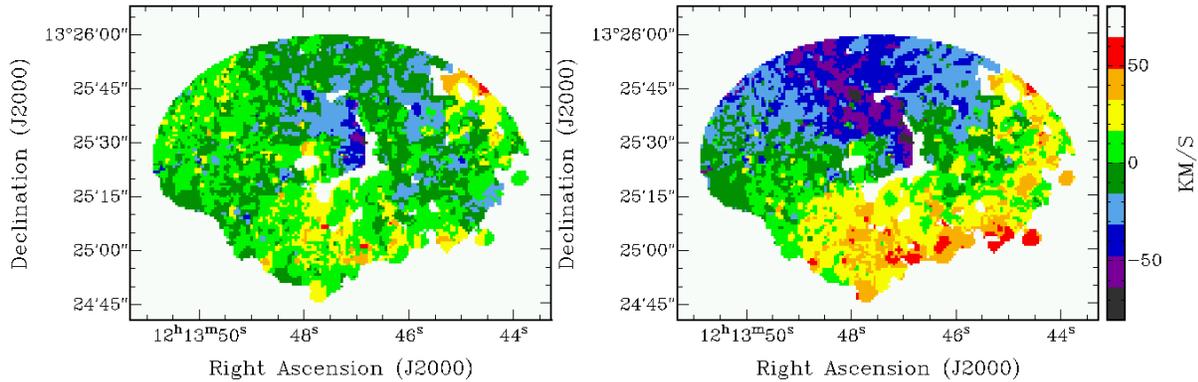}
\caption[]{Residual velocity maps of NGC 4189. (Left) Using the kinematical \pa\ of 251\degr\ and the 
tilted-ring model rotation curve (Fig.~\ref{vrotn4189} bottom).  (Right) Using the photometric \pa\ of 275\degr\ and the
smooth  model of the long-slit data  (Fig.~\ref{vrotn4189} top). The velocity range and 
the colour-scale are the same for the 2 maps.}
\label{resn4189}
\end{figure*}

As we wish to compare our derived rotation curves with results from earlier optical studies 
(Chincarini \& de Souza 1985; Sperandio et al. 1995; Rubin et al. 1999), 
it is first needed to ensure whether the FP measurements agree  with the long-slit data, 
at least along the  \textit{photometric} major axis. 
 An example  is illustrated in Fig.~\ref{vrotn4189} with  the galaxy NGC 4189. 
Both long-slit and FP  rotation curves extracted along the major axis with a constant \pa\ of 265\degr\ agree very well. 
However the average kinematical \pa\ (251\degr) clearly differs from the photometric one. 
 As a consequence, the rotation curve of the galaxy derived with the tilted-ring model differs from 
 the long-slit one (bottom panel of Fig.~\ref{vrotn4189}). 
To check that our result is a better representation of the kinematical pattern than the long-slit one, 
it is displayed on Fig.~\ref{resn4189} the residual velocities 
obtained by subtracting from the observation two models. A first velocity  model is based on the tilted ring fitting to the FP velocity field (left panel) 
and another model is based on the smooth long-slit rotation curve with a \pa\ value fixed to the photometric one (right panel).  
Notice that this smooth model does not take into account the last two points of the long-slit rotation 
curve and that it does not pass through the zero velocity. This does not change the following result. 
The figure shows residual velocities more homogeneously distributed in the disc for our model, with a dispersion of 16 \kmps\ in the residuals, 
whereas there is an excess of residual velocities on both sides of the major axis for the long-slit model, with a worst  
dispersion of 28 \kmps\ in the residuals.  
This signature is the one expected in a residual map when the model uses an erroneous \pa\ (see Warner, Wright \& Baldwin 1973). 

As explained in \S\ref{compphk} the photometry and the kinematics are in a relative good agreement, implying that 
long-slit data  should only occasionally give incorrect results, like with NGC 4189 when a strong misalignment is observed. 
This claim will be tested within a future work for all the Virgo sample. 


\subsection{Non-axisymmetry of gaseous velocity fields}
\label{asymmetry}
Galaxies in clusters are sensitive to the environmental effects. 
Their structures (gas, stars, dark matter halo) can be perturbed  by the cluster tidal field, 
gravitational encounters with other galaxies, galaxy mergers,  
accretion of gas, etc... Such external events can trigger internal  perturbations in a stellar 
disc like bars or oval distorsions (e.g. Noguchi 1987; Bournaud \& Combes 2002), spirals,  warps 
(Huang \& Carlberg 1997), polar rings (Bournaud \& Combes 2003) or lopsidedness (Bournaud et al. 2005). 
Ram pressure stripping is furthermore efficient in creating lopsided gaseous discs and off-plane 
annular-like gas structures (e.g. Schulz \& Struck 2001; Vollmer et al. 2001).
All these perturbations are likely responsible for the formation of the kinematical asymmetries 
observed in  our \ha\  velocity fields (Appendixes~\ref{comments} and~\ref{velmap}) and in 
other \hi\ or \ha\ data (Swaters et al. 1999; Beauvais \& Bothun 1999, 2001;  Verheijen \& Sancisi 2001), as well as  
in their optical rotation curves (Rubin et al. 1999; Dale et al. 2001). 

 Franx, van Gorkom \& de Zeeuw (1994), Schoenmakers et al. (1997) 
 or Wong et al. (2004) made an analysis of \hi\ and CO velocity fields by means of Fourier 
 coefficients  in order to study the non-axisymmetry of the gas kinematics. A similar approach has been used 
 with stellar kinematics of early-type galaxies by means of kinemetry (Copin et al. 2001; Krajnovi\'c et al. 2005). The technique decomposes the 
  line-of-sight velocity along each ring of the tilted-ring model into harmonic terms $c_m$ and $s_m$ of order $m$ following 
$$v_{\rm obs} = c_0 + \sum_m{c_m \cos(m\Psi) + s_m \sin(m\Psi)}$$ where $\Psi$ corresponds to the angle in the plane 
of the ring, the coefficient  $c_0$  to the systemic velocity of the ring, the first order term  $c_1$ 
to more or less  the rotation curve  and all other terms  to non-circular motions (axisymmetric and non-axisymmetric).  
Schoenmakers et al. (1997) showed that a galaxy perturbed by a distortion of the potential of harmonic order $m$  
exhibits terms of order $m-1$ and $m+1$ in its velocity field. 

 \begin{figure}
\centering
   \includegraphics[width=8cm]{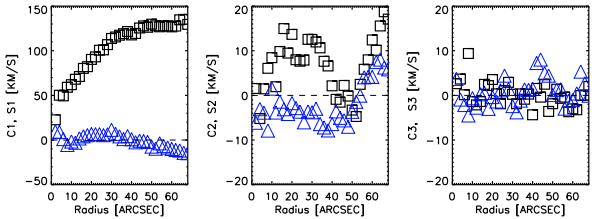}
\caption[]{Harmonic decomposition of the velocity field of NGC 4298. Squares (triangles) correspond to the cosine (sine, resp.) 
terms.}
\label{exn4298}
\end{figure}

 This technique has also been applied to ionized gas  velocity fields (Krajnovi\'c et al. 2005; Simon et al. 2005) 
 and  reveals itself here to be successful  in detecting the signatures of non-axisymmetry of the 2D \ha\ kinematics, as seen in 
Figure~\ref{exn4298}.  This diagram displays the Fourier coefficients as a function of radius after an harmonic 
expansion up to an order of $m=3$ of the velocity field of NGC 4298, using fixed and constant orientation parameters (given
in Tab.~\ref{kinpar}), kinematical centre position and systemic velocity.
The $s_1$ term is not null and the $s_3$ term appears to slightly oscillate around zero. There is no bar nor warp 
in NGC 4298 so that these variations could be due to the presence of a weak $m=2$ spiral pattern. 
The most striking feature is the significant variations of the $m=2$ kinematical terms. 
This can be caused by  $m=1$ or $m=3$ modes. This latter hypothesis can be 
rejected because the \ha\ morphology of NGC 4298 does not show signatures of $m=3$ structures like 
e.g. 3 spiral  arms. There is no direct evidence of a ``one spiral arm'' structure in NGC 4298 so that the $m=1$ mode 
could be caused by a kinematic lopsidedness.  The presence of the galaxy NGC 4302, companion 
of NGC 4298 in the sky plane but also in radial velocity, could be at the origin of the observed asymmetries. 
Beauvais \& Bothun (1999) also observed drifting centres in their FP observations. 
While a large number of galaxies appear to exhibit a \emph{morphological} lopsidedness, as revelead by 
 galaxy imagery (Bournaud et al. 2005) or \hi\ data (Swaters et al. 1999; Noordermeer et al. 2005), 
evidence of \emph{kinematical} lopsidedness such as the one probably detected here in NGC 4298 seem to be  
rare in the WHISP sample of Noordermeer et al. (2005).

A forthcoming work will apply the Fourier analysis to the Virgo cluster sample 
(as well as velocity maps of field galaxies) and will investigate how the kinematical perturbances 
can be related to the cluster environment.

%
%
%
%
\section{Conclusion}
\label{conclusions}
This article has  presented  Fabry-Perot interferometry observations of the ionized gas 
 for a sample of 30  spiral and irregular galaxies of the Virgo cluster. 
 This is the first time that a sample of \ha\ integrated emission-line maps and 
velocity fields are collected in order to study the environmental effects on the 
two-dimensional, high-resolution kinematics of cluster galaxies. 
Two new  improvements in the reduction of the FP data 
consist in  a better sky subtraction  and  an adaptive binning. Both these 
procedures allow  to achieve an optimal signal-to-noise in the field-of-view  
and a better detection of  the diffuse ionized gas in interarm or external regions.

The galaxies exhibit morphological irregularities like nuclear spiral, nuclear and inner pseudo-ring, 
off-plane diffuse filaments and arc-like structure, truncated disc, lopsidedness, bar and spiral arms. 
The velocity fields exhibit  typical kinematical perturbations like  
streaming motions along spiral arms, twist of the major axis, Z-shape of velocities 
due to the presence of a bar, decoupled nuclear spiral or misalignment between photmetric and kinematical major axes. 
Future papers from  this series will present the rotation curves and mass models of the galaxies 
and investigate whether the asymmetries observed in the velocity fields have a relation with such a dense environment.

%
%
%
\section*{Acknowledgements} 
We are grateful to the anonymous referee for valuable comments. 
We thank all the  staffs of the observatories which helped us during the different runs. We are very grateful 
to B. Malenfant and G. Turcotte at the Observatoire du mont M\'egantic,  
to  O. Daigle, J.-L. Gach, M.-M. de Denus-Baillargeon, M.-H. Nicol and D. Naudet
 for their help during the observations or the preparation of the article. 
 L. Chemin thanks F. Combes and F. Bournaud for insightful conversations. 
 C. Balkowski thanks the PPF program at the Observatoire de Paris for financial support. 
This work was partly funded by   the grant Regroupement Strat\'egique -
Observatoire du mont M\'egantic of FQRNT (Qu\'ebec) and by the Minist\`ere de l'\'Education Nationale, de la Recherche 
et de la Technologie (France). 
The \fantomm\ project has been carried out by the Laboratoire d'Astrophysique Exp\'erimentale (LAE) of the Universit\'e de Montr\'eal 
using grant from the Canadian Foundation for Innovation and the Minist\'ere de l'\'Education du Qu\'ebec.
We made use of the HyperLeda (http://leda.univ-lyon1.fr) and Nasa/IPAC Extragalactic Database (http://nedwww.ipac.caltech.edu) databases.
This research has made use of the GOLDMine database and products : 
Galaxy On Line Database Milano Network (http://goldmine.mib.infn.it). 
This publication makes use of data products from the Two Micron All Sky Survey, which is a 
joint project of the University of Massachusetts and the Infrared Processing and Analysis 
Center/California Institute of Technology, funded by the National 
Aeronautics and Space Administration and the National Science Foundation.
%
%
%

%
%
\appendix

\section{The kinematical \ha\ catalog  of Virgo galaxies}
\label{kinematics}

\subsection{Comments for individual galaxies}

\label{comments}

\begin{description}
\item[\bf NGC 4189:]
The velocity field  of this barred multi-armed galaxy is slightly perturbed.
 The position-velocity diagram reveals that the \ha\ peak in the nucleus has an almost constant velocity,
which signature was already observed in the long-slit data (Rubin et al. 1999).  The kinematical and photometric  major axes 
are not aligned.

\item[\bf NGC 4212:]
The galaxy presents a mildly perturbed velocity field in the northern and western regions.
The optical and  near-infrared images clearly reveal an important dust obscuration effect in the galaxy.
This may disturb the gaseous kinematics of NGC 4212, as particularly
seen in the western region.
The long-slit data showed velocities down to $\sim -240$ \kmps\ (Rubin et al. 1999), which are not observed here.
 As in the CO data (Sofue et al. 2003a), the core exhibits a steeper velocity gradient than for the rest of the disc.
 This could be due to the presence of a hidden bar.  A careful analysis of the NIR image by an isophotal ellispses fitting indeed seems to reveal 
 the presence of an elongated inner structure (see also Buta et al. 2005). 
  
\item[\bf NGC 4237:]
The velocity field shows no real signs of perturbations, except for a clump in the receding half to the South
(R.A. $\sim$ \coa{12}{17}{12.53}, Dec. $\sim$ \cod{+15}{18}{58}), which exhibits a
lower velocity ($\sim$ 900 \kmps) than its surroundings ($\sim 950$ \kmps). 
Streamings of gas are also observed.

\item[\bf NGC 4254:]
The prominent spiral structure is asymmetric in this well known-galaxy, as well as its \ha\ velocity field.
Streaming motions are observed along the spiral arms.
Our observation of NGC 4254 is in good agreement with the \ha\ and \hi\ velocity fields presented in Phookun et al. (1993).
Notice a region at R.A. $\sim$ \coa{12}{18}{57.56}, Dec. $\sim$ \cod{+14}{26}{55}, $v_{\rm obs} \sim 2535$ \kmps) having a higher velocity
than its surroundings ($\sim 2500$ \kmps). 
CO data show a steep velocity rise in the innermost arcseconds (80 \kmps).  
It could be due to a small bar detected by Kranz et al. (2001).

\item[\bf NGC 4294:]
With NGC 4299, NGC 4294 belongs to an apparent pair. Streaming motions are seen along the spiral arms. The
receding half appears slightly pertubed, with \hii\ regions having lower radial velocities
than the local diffuse gas  (e.g. at R.A. $\sim$ \coa{12}{21}{16.92}, Dec. $\sim$ \cod{+11}{31}{01}).

\item[\bf NGC 4298:]
This apparently flocculent galaxy (Elmegreen et al. 1999) has a companion NGC 4302 only distant by  \am{2}{3} (or 10.7 kpc projected) and with nearly the same
systemic velocity ($v_{\rm sys}^{N4302}=1149$ \kmps, Binggeli et al. 1985). According to Koopmann \& Kenney (2004), the truncated gas disc is due to 
ram pressure stripping and the stellar asymmetry to a tidal interaction with a companion. 
The velocity field  appears however mildly perturbed. Notice that the southern arm presents velocities 
more redshifted than the nearby parts of the disc at the same azimuthal angles. This perhaps indicates 
streaming motions along it or, less probably, a locally warped arm. A steep velocity rise is detected in 
the PV-diagram in the core of the galaxy.

\item[\bf IC 3225:]
This magellanic-type galaxy has an enhanced star formation on the NE, approaching half of its disc 
whereas very little ionized gas is detected on the SW half. 
The $B-$band distribution shows an elongated stellar disc to the SW whereas 
the NIR image has a truncated disc in this direction. 
As seen in the position-velocity diagram, the velocity smoothly increases with radius.
The velocity field thus mainly displays a solid body rotating pattern. This is typical of highly inclined optically thick discs,
 which projection effects prevents from observing the true rotation (Goad \& Roberts 1981; Giovanelli \& Haynes 2002).

\item[\bf NGC 4321:]
The well-known galaxy NGC 4321 (M100) has often been mapped with Fabry-Perot interferometry (Arsenault et al. 1990;
Knapen et al. 1995; Canzian \& Allen 1997; Hernandez et al. 2005b). Evidence for three different 
pattern speeds (nuclear structure, large-scale bar, spiral)  has been shown using the Tremaine-Weinberg method on this rather 
regular velocity field (Hernandez et al. 2005b), which also show signs of streaming motions (Knapen et al. 2000). 
The core of the galaxy has a nuclear ring-like structure.

\item[\bf NGC 4351:]
The galaxy is slowly rotating. Its has a peculiar gaseous morphology,  showing   an arm-like structure  to the North.
The stellar distribution appears lopsided, also with an arm-like structure in the northern and western regions of the galaxy.
 The stellar asymmetry could result from a past tidal interaction, or even a merger. 
This galaxy is also a very good candidate for an ongoing RPS (Koopmann \& Kenney 2004).

\item[\bf NGC 4405:]
The \ha\ distribution is truncated and the velocity map  displays a solid-body rotating pattern.
This is understandable in a framework where only the central regions lying in the rising part of a rotation curve are observed.

\item[\bf NGC 4438:]
 Notice the off-plane filamentary morphology of the ionized gas
to the  West and South of the galaxy disc, which was already seen in Kenney et al. (1995) and Kenney \& Yale (2002). 
No \hi\ was detected in the disc (Cayatte et al. 1990). The ionized gas is the best tracer of the complex morphology 
and kinematics of the galaxy, which is presented in detail in Chemin et al. (2005).

\item[\bf NGC 4450:]
Like in the \hi\ data of Cayatte et al. (1990), this anemic galaxy has a patchy \ha\ distribution. The \ha\ spiral    
appears highly wound around the centre. 
The velocity field appears significantly dispersed in the galaxy inner 
parts as well as in the outer ones but to a lesser extent. 
The nucleus displays a steep gradient ($\sim 200$ \kmps), which result was already seen in long slit data (Rubin et al. 1999).
Little CO gas is observed in NGC 4450 (Helfer et al. 2003).

\item[\bf NGC 4457:]
The  distribution and the kinematics of this galaxy are very peculiar.  There is little ionized gas in the  receding side,
which is traced by only few star formation regions while the approaching side consists in a single \ha\ armlet.
 The core region surprisingly exhibits the most blue-shifted velocities observed in the galaxy. 

\item[\bf NGC 4498:]
The emission is mostly located in \hii\ regions along the stellar bar-like structure. The kinematics of NGC 4498 is asymmetric: the approaching
half is typical of a constant velocity rotational pattern while the receding side has solid-body rotation.  

\item[\bf NGC 4501:]
The \ha\  velocity map of NGC 4501 (M88) appears regular with the exception 
of non-circular motions associated to the spiral structure, as particularly 
seen along the minor axis. NGC 4501 exhibits one of the largest velocity range among all the galaxies presented here,
indicating a very massive object. Notice that part of the NW \ha\ spiral arm is 
missing  in the approaching half of the disc, which is probably
due to a dust obscuration effect and to a low transmission of the interference filter.
The CO kinematics in the $\pm 4$\arcsec\ region around the photometric 
centre is complex, as revealed by double-peaks in the molecular gas emission (Onodera et al. 2004). 
The \ha\ PV-diagram also shows a complex central kinematics. 
Cayatte et al. (1990) showed compressed contours of \hi\ distribution to the SE.

\item[\bf NGC 4519:]
The galaxy morphology is perturbed, as well as the velocity field where streaming motions
are revelead along the multiple spiral arms. Notice the strong 
 Z-shape of the velocities due to the presence of the bar. This galaxy has a  
misalignment between  photometric and kinematical major axes. 

\item[\bf NGC 4532:]
 The velocity map is very perturbed. It displays a strong Z-shape kinematical pattern,
 which has already been observed in the \hi\ velocity field (Hoffman et al. 1999).
A systemic velocity component ($\sim 2010$ \kmps) is observed   along the north-western
edge of the \ha\ distribution. 
NGC 4532 has the highest \ha\ equivalent width in the sample of Koopmann \& Kenney (2004).
It is not surprising while regarding its perturbed stellar and gaseous morphologies,
with prominent star forming regions (like e.g. the off-plane one at R.A. $\sim$ \coa{12}{34}{21.02}, Dec. $\sim$ \cod{+06}{27}{38}) and stellar extensions.
All the perturbations may result from a tidal interaction and/or gas accretion
(Hoffman et al. 1999). Notice that an optical reflection of the galaxy was masked, giving  artificial sharp edges 
to the shape of the galaxy  in its eastern parts.

\item[\bf NGC 4535:]
This barred galaxy has a low inclination and an irregular morphology, as mostly seen in the $B$-band and \ha\ images,
 with several spiral-like structures arising to the North and South in addition to the main spiral pattern.
 The velocity field is perturbed in the northern \ha\ loop and in the southern arm-like extension.
 Streaming motions are also observed along the arms. 
  The PV diagram   shows a steep velocity rise associated with the bright \ha\ 
 core. The central regions of NGC 4535   have no neutral gas (Cayatte et al. 1990).

\item[\bf NGC 4536:] This large barred galaxy displays two main prominent spiral arms and
additional spiral-like structures. The only kinematical perturbations
are seen in the Z-shape of the velocities in the central parts and streaming motions along the arms.
As in the CO data (Sofue et al. 2003a), a steep velocity gradient is observed in the \ha\ data in
 the galaxy core.

\item[\bf NGC 4548:] The observation  has a very low initial $S/N$ which causes
the bin sizes to be large during the Voronoi tessellation. The disc rotation is nonetheless seen, with a
velocity gradient of $\sim$ 200 \kmps, which result is in agreement with that deduced from long-slit spectroscopy (Rubin et al. 1999).

\item[\bf NGC 4567:]  This galaxy is part of a pair with NGC 4568. The morphology and the velocity field do
not show signs of disturbance.

\item[\bf NGC 4568:] 
Here again, no clear signs of velocity disturbance are observed, except perhaps in the external regions of the disc.
 Koopmann \& Kenney (2004) claim that
this galaxy pair likely experiences RPS and a tidal interaction.
It is however not obvious that a tidal interaction is (or has) occuring (occured)
while regarding the very mildly perturbed kinematics of the pair. 

\item[\bf IC 3583:] The galaxy presents irregular stellar and gaseous morphologies. 
Tsch\"{o}ke et al. (2001) revealed the presence of a faint, narrow spur of star formation regions to the South, pointing 
toward NGC 4569. This spur is also seen in the FP field.
A  weak rotation is evidenced in the  perturbed  velocity map ($\sim 100$ \kmps\ in projection).

\item[\bf NGC 4569:] The velocity field of the galaxy is perturbed, showing gas streamings  along the arm-like
structures in the \ha\ (severely) truncated disc. An off-plane string of \hii\ regions is observed to the West of
the disc. It exhibits a velocity gradient that follows the rotation of the disc.
This gaseous structure has already been seen in deep \ha\ image (Tsch\"{o}ke et al. 2001) and in \hi\ data (Vollmer et al. 2004).
The giant diffuse gaseous region to the West of the disc (Tsch\"{o}ke et al. 2001) is not detected in the FP observation, 
which is probably due to a too low $S/N$ or to a likely cut-off in the transmission of the interference filter.
Vollmer et al. (2004) proposed a ram pressure stripping model to explain the \hi\ asymmetry.

\item[\bf NGC 4571:]
This almost face-on galaxy has a very patchy distribution of \hii\ regions. The interference filter
unfortunately cuts part of the emission in the NE region.  The 21-cm data show an \hi\ distribution
 well confined in the optical disc boundaries (van der Hulst et al. 1987).
 The \ha\ kinematics  is in agreement with the \hi\ data and presents a slightly perturbed velocity field.

\item[\bf NGC 4579:]
The FP data presents an \ha\ kinematics of NGC 4579 (M58) which is perturbed by the streaming motions along the large-scale
spiral arms and by a rather large velocity dispersion in the receding half (by comparison with the approaching side).
The circumnuclear region is particularity interesting because it
exhibits a bright ``loop'' structure within a radius of $\sim$ 1.5 kpc (Keel 1983; Pogge 1989; Gonz\'alez Delgado \& Perez 1996)
and the nucleus is known to have an extremely broad \ha\ profile with a full-width at zero intensity of up to 18 000 \kmps\ (Barth et al. 2001).
The so-called ``loop'' is actually an asymmetric nuclear spiral which displays a very large velocity gradient ($\sim 500$ \kmps). 
The kinematical major-axis of the nuclear spiral is almost perpendicular to that of the large-scale spiral structure.
The high-resolution data of the molecular gas component showed CO emission concentrated within the core (Helfer et al. 2003;  
Sofue et al. 2003a) and the neutral hydrogen emission was lying in an annulus-like structure (Cayatte et al. 1990) 
which is actually the large scale spiral arms. 

\item[\bf NGC 4580:]
 This \ha\ truncated disc displays  little kinematical irregularities and a small
velocity gradient. Though the \ha\ covers the innermost parts of the stellar distribution like in NGC 4405, a
constant velocity seems to be reached along the major-axis, which was not seen in NGC 4405. 
The \ha\ morphology is in the form of a pseudo-ring. No ionized gas is detected in the spiral arms that can be seen in the 
broad-band images. 

\item[\bf NGC 4639:]
This barred galaxy exhibits a remarkable inner ring (Evans et al. 1996; Gonz\'alez Delgado et al. 1997).
The outer gaseous distribution is furthermore reminiscent of an outer pseudo-ring.
This outer structure is visible in the $B$-band image but hardly in the NIR image, which principally shows the bar and the bulge.
The velocity field is rather regular, consistent with a flat circular velocity pattern of the disc. It also shows streaming 
motions in the ring. A steep velocity rise is seen in the galaxy core, as was also observed in Rubin et al. (1999). 

\item[\bf NGC 4654:]
The stellar and gaseous morphologies are very asymmetric. The \ha\ velocity field is strongly perturbed.
Alternating patterns of (decreasing and increasing) solid-body and constant velocity are observed through the field,
which is well illustrated by the PV diagram. In the approaching half, the transition region between the central solid body part
and the outer decrease occuring at an offset of $-1\arcmin$ leaves a remarkable imprint in the velocity field. 
 In addition, streaming motions are evidenced along the spiral arms. 
 A solid-body rotation is partly associated with the very thin and elongated \ha\ bar-like structure in the central parts.
According to Vollmer (2003), a scenario mixing effects of a tidal interaction   
with NGC 4639 $\sim 500$ Myrs ago and RPS could be responsible of the perturbed stellar and gaseous distributions of NGC 4654. 
There is a large difference of distance between the two galaxies, $\sim 22$ Mpc and $\sim 25$ Mpc for NGC 4639 
as estimated from the Tully-Fisher relation (Yasuda et al. 1997) and the Cepheids method (Saha et al. 1997) respectively, 
and $\sim 14$ Mpc for NGC 4654 (Tully-Fisher relation, Yasuda et al. 1997). This inconsistency of distance points out that  
 the Tully-Fisher distance of NGC 4654 must be highly erroneous if the Vollmer's scenario is valid.  

\item[\bf NGC 4689:] As for NGC 4571, the \ha\ emission is very patchy. 
The velocity field appears slightly perturbed, showing evident signs of streamings  
associated with the pseudo-spiral arms. The \ha\ kinematics is in good agreement 
with the \hi\  (Cayatte et al. 1990) and  CO (Sofue et al. 2003a) data.
\end{description}

\subsection{Observational data}
\label{velmap} 

 \LARGE\textbf{Due to the restricted size allowed for the astro-ph files, only the example of NGC 4254 is presented here. 
All other 29 images can be previewed on http://www.astro.umontreal.ca/fantomm/virgo~. Full ps (36 Mb) and pdf (3Mb) resolution 
versions with all images are available on ftp://ftp.astro.umontreal.ca/outgoing/chemin}

\begin{figure*}
\begin{center}
\Huge \textbf{NGC 4254}
    \includegraphics[width=\textwidth]{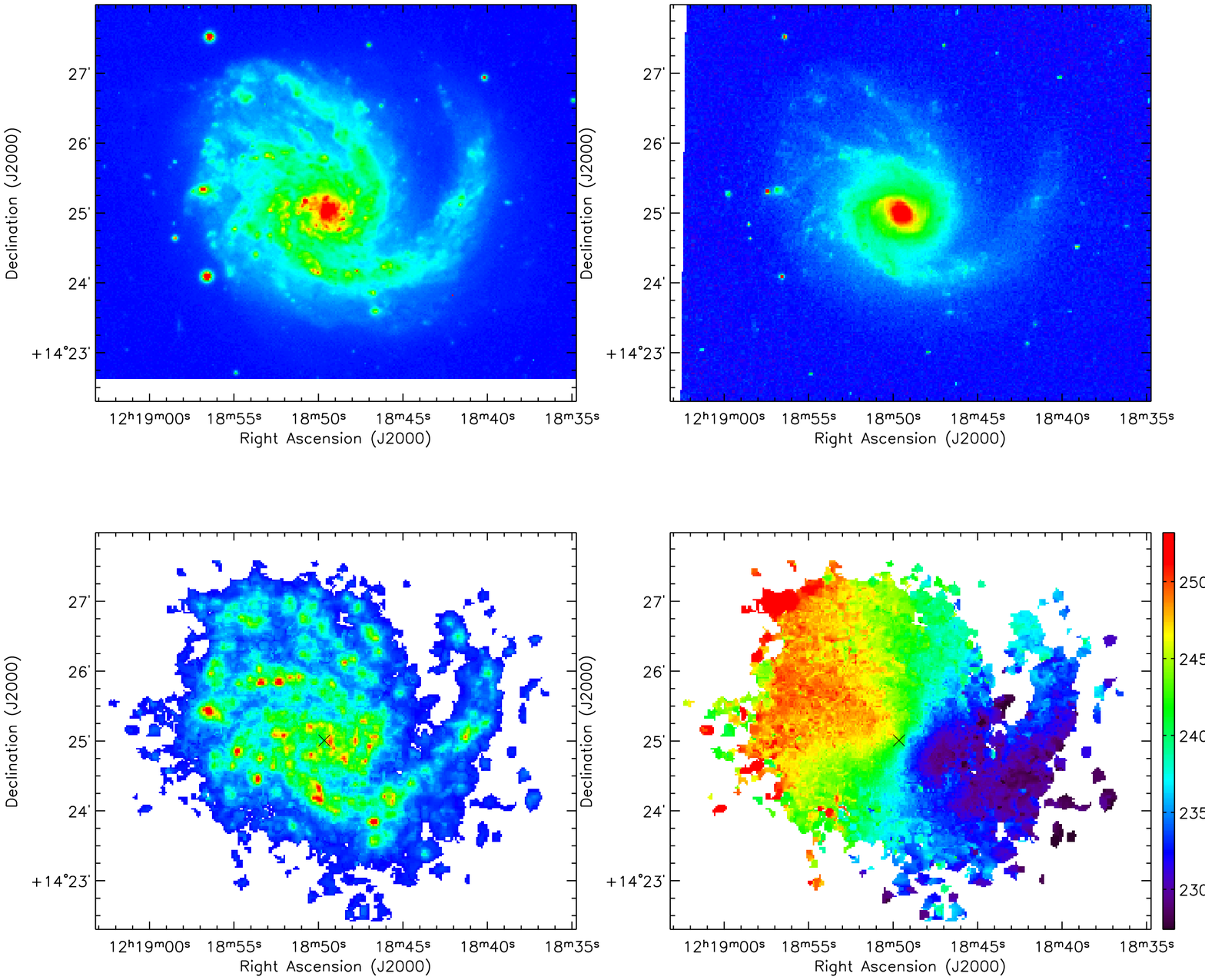}\vspace{-1.9cm}
  \includegraphics[height=16cm,angle=270]{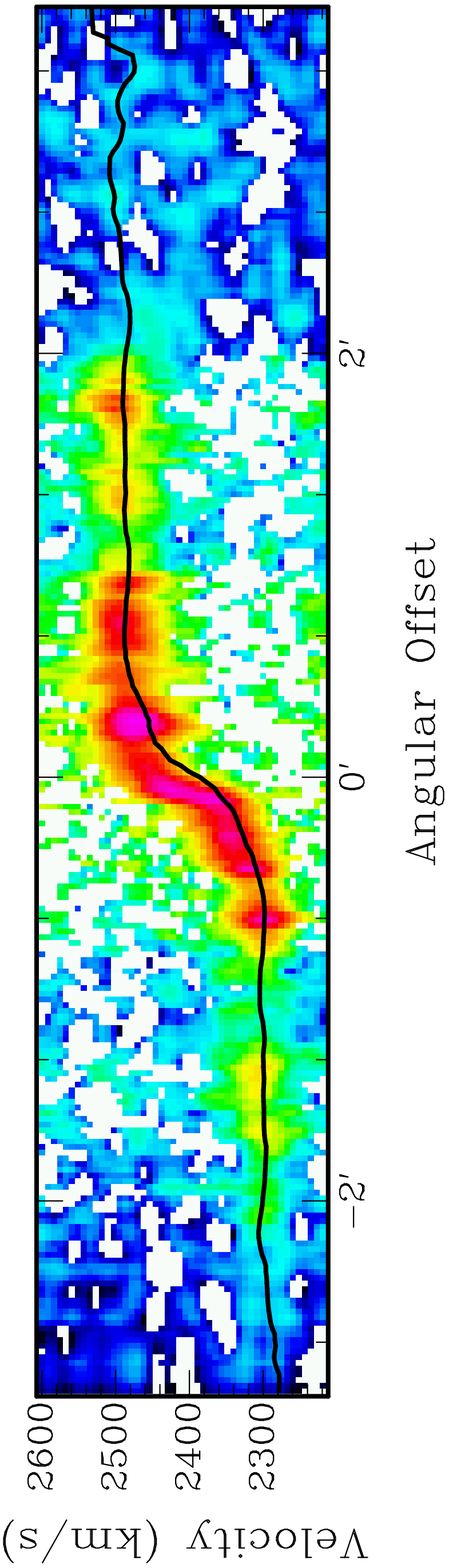}
\end{center}
\vspace*{-0.2cm}
\caption[]{Presentation of the \ha\ maps of the 30 Virgo galaxies. NGC 4254 (VCC 0307).
  Top : GOLDMine $B$-band (Left) and $H$-band (Right) images. Bottom : FP \ha\ integrated map (Left) and
  \ha\ velocity field (Right).  All three images of integrated intensities are displayed using a square root stretch with increasing
  intensity levels from blue to red colours.  In the \ha\ flux and velocity maps, a cross gives the position of the photometric centre.
    The velocity field is displayed using a linear scale and are in \kmps. The PV diagram is displayed using a logarithmic stretch. 
  A solid line represents the derived rotation curve of the galaxy (see section~\ref{results}).  1\arcmin\ corresponds to
  4.7 kpc.}
\label{fg1}
\end{figure*}

\end{document}